\documentclass[sigconf]{acmart}
\usepackage{xspace}
\usepackage{algorithm}
\usepackage{algpseudocode}
\usepackage{graphicx}
\usepackage{subcaption}
\usepackage{xspace}
\usepackage{adjustbox}
\usepackage{enumitem}
\usepackage{float}

\usepackage{multirow}
\usepackage{graphicx}
\usepackage{arydshln}  
\usepackage{balance}

\algrenewcommand\algorithmicrequire{\textbf{Input:}}
\algrenewcommand\algorithmicensure{\textbf{Output:}}
\AtBeginDocument{%
  }

\copyrightyear{2026}
\acmYear{2026}
\setcopyright{cc}
\setcctype{by}
\acmConference[WWW '26]{Proceedings of the ACM Web Conference 2026}{April 13--17, 2026}{Dubai, United Arab Emirates}
\acmBooktitle{Proceedings of the ACM Web Conference 2026 (WWW '26), April 13--17, 2026, Dubai, United Arab Emirates}
\acmPrice{}
\acmDOI{10.1145/3774904.3792684}
\acmISBN{979-8-4007-2307-0/2026/04}

\newcommand{\method}{\textsc{RoE}\xspace }

\begin{document}

\title{Reasoning by Exploration: A Unified Approach to Retrieval and Generation over Graphs}


\author{Haoyu Han}
\orcid{0000-0002-2529-6042}
\affiliation{
  \institution{Michigan State University}
  \city{East Lansing}
\state{MI}
  \country{USA}
}
\email{hanhaoy1@msu.edu}

\author{Kai Guo}
\authornote{Corresponding Authors.}
\orcid{0000-0002-3841-8862}
\affiliation{
  \institution{Michigan State University}
    \city{East Lansing}
\state{MI}
  \country{USA}
}
\email{guokai1@msu.edu}

\author{Harry Shomer}
\orcid{0000-0001-5081-1870}
\affiliation{
  \institution{University of Texas at Arlington}
    \city{Arlington}
\state{TX}
  \country{USA}
}
\email{harry.shomer@uta.edu}

\author{Yu Wang}
\orcid{0000-0001-6908-508X}
\affiliation{
  \institution{University of Oregon}
      \city{Eugene}
\state{OR}
  \country{USA}
}
\email{yuwang@uoregon.edu}
\author{Yucheng Chu}
\orcid{0009-0004-0072-5455}
\affiliation{
  \institution{Michigan State University}
      \city{East Lansing}
\state{MI}
  \country{USA}
}
\email{chuyuch2@msu.edu}
\author{Hang Li}
\orcid{0000-0002-3464-3245}
\affiliation{
  \institution{Michigan State University}
        \city{East Lansing}
\state{MI}
  \country{USA}
}
\email{lihang4@msu.edu}

\author{Li Ma}
\authornotemark[1]
\orcid{0000-0002-5712-2143}
\affiliation{
  \institution{Michigan State University}
        \city{East Lansing}
\state{MI}
  \country{USA}
}
\email{mali13@msu.edu}

\author{Jiliang Tang}
\orcid{0000-0001-7125-3898}
\affiliation{
  \institution{Michigan State University}
          \city{East Lansing}
\state{MI}
  \country{USA}
}
\email{tangjili@msu.edu}

\renewcommand{\shortauthors}{Haoyu Han et al.}

\begin{abstract}
Reasoning over structured graphs remains a fundamental challenge for Large Language Models (LLMs), particularly when scaling to large graphs. Existing approaches typically follow the retrieval-augmented generation (RAG) paradigm: first retrieving subgraphs relevant to the query and then generating answers conditioned on the retrieved subgraphs. However, such two-phase pipelines often struggle to faithfully incorporate graph structure, since the generation process is ultimately constrained by the quality and completeness of the retrieved subgraph. Although many advanced retrievers have been proposed recently to mitigate this issue, they are usually tailored to the training graphs and generalize poorly to unseen graphs, which limits their practical applicability. In this work, we propose Reasoning by Exploration (\method), a novel approach that unifies retrieval and generation by framing reasoning over graphs as a process of graph exploration. At each step, the LLM selects candidate nodes and edges to explore, gradually constructing reasoning paths and generating answers along the way. To enable effective exploration, RoE is trained in two stages: supervised fine-tuning (SFT) on gold reasoning paths, followed by reinforcement learning (RL) to enhance exploration effectiveness and generalization. Experiments on benchmark datasets demonstrate that RoE achieves substantial overall improvements over baselines, while also generalizing effectively to unseen graphs. The code can be found at \url{https://github.com/haoyuhan1/RoE}.
\end{abstract}




\maketitle

\section{Introduction}
\label{sec:intro}
Large language models (LLMs)~\citep{chang2024survey, thirunavukarasu2023large, kasneci2023chatgpt} have demonstrated impressive abilities across a wide range of natural language processing tasks. To further enhance their reliability and factuality, Retrieval-Augmented Generation (RAG)~\cite{lewis2020retrieval, gao2023retrieval, jiang2023active, wang2024knowledge} has become a widely adopted paradigm. By retrieving external knowledge before generating responses, RAG has achieved strong performance in diverse domains, such as healthcare~\citep{xu2024ram}, law~\citep{wiratunga2024cbr}, finance~\citep{zhang2023enhancing}, and education~\citep{chu2025enhancing}. However, most existing RAG systems operate over unstructured text, retrieving passages through lexical or semantic similarity, and therefore do not fully leverage the relational structure inherent in many forms of knowledge.

Graphs provide a natural representation for structured knowledge, capturing rich relational dependencies among entities. They are widely used in domains such as knowledge bases, social networks, and scientific applications including biology and chemistry~\citep{wu2020comprehensive, ma2021deep, wang2020traffic}. Recently, Graph Retrieval-Augmented Generation (GraphRAG) has emerged as a promising way to integrate such graph-structured knowledge into LLMs~\citep{han2024retrieval, peng2024graph, lei-etal-2025-mixture}. There are mainly two steps in GraphRAG: first, a retriever selects query-relevant content from the entire graph, where the retrieved units may be nodes, paths, or subgraphs; 
and then, a generator, usually an LLM, produces answers conditioned on the retrieved subgraphs. This framework has been successfully applied to tasks such as knowledge graph question answering (KGQA)~\citep{yasunaga2021qa, yasunaga2022deep, zhang2022greaselm}, where often requires multi-step reasoning over the graphs.

While efficient, this two-phase pipeline may constrain reasoning quality due to the disjoint optimization of retriever and generator. 
{\bf First}, the retriever often ignores the specific needs of the generator, which can lead to missing critical evidence or introducing irrelevant noise. For example, consider the multi-hop question “Who are the nephews of Winston Churchill?”. This query may have multiple answers and can be reached through multiple reasoning paths, such as ``Churchill $\rightarrow$ sibling $\rightarrow$ child" or ``Churchill $\rightarrow$ parent $\rightarrow$ child $\rightarrow$ child". Heuristic-based retrievers~\cite{yasunaga2022deep, feng2020scalable} (e.g., retrieving fixed $k$-hop subgraphs) or GNN-based retrievers~\cite{mavromatis2024gnn, liu2024dual} may include a large amount of irrelevant information about Churchill, overwhelming the generator with noise~\cite{guo2025empowering}.
In contrast, LLM-based retrievers~\cite{luo2023reasoning, wu2023retrieve} may generate only a limited set of reasoning paths, leading to missing valid entities and incomplete answers, as shown in Section~\ref{sec:pre}. {\bf Second}, retrievers or generators trained on specific graphs or datasets usually generalize poorly to unseen graphs and question types as shown in Section~\ref{sec:pre}. For instance, a path-based retriever tuned on a movie knowledge base may learn to generate reasoning paths primarily around relations such as ``actor $\rightarrow$ film $\rightarrow$ director", but it may struggle to generalize when applied to graphs in other domains with very different relational structures, such as biomedical graphs involving ``gene $\rightarrow$ protein $\rightarrow$ disease" paths. This dependence leads to sharp performance drops when deployed on unseen graphs, limiting the robustness and real-world applicability of current GraphRAGs.

To address these limitations, we propose Reasoning by Exploration (\method), a unified framework that leverages the reasoning ability of LLMs to perform retrieval and generation simultaneously within the same model. Instead of separating the two stages, \method frames reasoning as a step-by-step graph exploration process, where the model incrementally expands nodes and edges while constructing reasoning paths and generating answers in a unified manner. In this way, \method has the access to the whole graph. This design directly mitigates the problem of missing or noisy retrieval, since the model actively decides which parts of the whole graph to explore based on its evolving reasoning process, instead of relying on a static retriever. However, it faces two main challenges. {\it First}, how can we equip LLMs, which pretrained primarily on unstructured text, with the ability to perform structured graph exploration? {\it Second}, how can we ensure that the model learns generalizable exploration strategies rather than memorizing specific graph patterns, so that it can adapt to unseen graphs? To tackle these challenges, we adopt a two-stage training strategy. In the first stage, we construct gold exploration trajectories and apply supervised fine-tuning (SFT) to teach the LLM to expand nodes and edges step by step. In the second stage, we employ reinforcement learning (RL) to further refine the exploration policy, encouraging the model to efficiently discover valid reasoning paths while improving its ability to generalize across diverse and unseen graphs. Together, this training paradigm enables \method to combine explicit supervision with adaptive learning signals, leading to more faithful reasoning and stronger generalization. 
We conduct extensive experiments on multiple multi-hop reasoning benchmarks, showing that \method consistently outperforms strong state-of-the-art baselines and exhibits robust generalization to unseen graphs and question types across different backbone sizes.


\section{Related Works}
\label{sec:related}
\subsection{GraphRAG}
GraphRAG~\cite{han2024retrieval, peng2024graph, lei2025mixture, han2025rag} aims to retrieve relevant information from external graphs to improve the performance and reduce the hallucination of LLMs. Compared to traditional RAG systems~\cite{lewis2020retrieval, gao2023retrieval} that operate on unstructured text, GraphRAG leverages the relational structure of graphs, where nodes represent entities and edges encode relations between nodes. Similar to RAG, GraphRAG generally consists of two main components: a retriever, which selects query-relevant graph content, and a generator, typically an LLM, which produces answers based on the retrieved subgraphs. To accommodate the graph structure, there are mainly three types of retrievers:

\textbf{Heuristic-based Retriever.} Heuristic-based retrievers typically begin by using similarity-based methods~\cite{wen2023mindmap, sanmartin2024kg} or entity and relation extraction techniques~\cite{al2020named, feng2020scalable} to identify seed nodes and edges that are relevant to the query. These seeds are then expanded by retrieving their $k$-hop neighbors~\cite{yasunaga2021qa, choudhary2023complex}, forming a candidate subgraph for reasoning. A representative example is G-Retriever~\cite{he2024g}, which first leverages embedding models to retrieve the most similar nodes and edges to the query, and then applies the Prize-Collecting Steiner Tree (PCST) algorithm to extract a compact subgraph.

\textbf{GNN-based Retriever.} GNN-based retrievers~\cite{liu2024dual, fang2024reano} typically train Graph Neural Networks (GNNs)~\cite{wu2020comprehensive, ma2021deep} to model retrieval as a node classification task, where the correct answer entities are labeled as positive and others as negative. For example, GNN-RAG~\cite{mavromatis2024gnn} first trains a GNN model and then retrieves the nodes with the highest probability scores as candidate answers, while also identifying the shortest paths connecting the question entities with these candidates to construct supporting evidence.

\textbf{LLM-based Retriever.} LLM-based retrievers leverage the reasoning ability of LLMs to generate retrieval plans instead of relying solely on graph embeddings or heuristics. For example, they may prompt an LLM to produce SQL queries over structured knowledge bases~\cite{jiang2024kg} or to explicitly plan multi-hop paths from the query entities to answers~\cite{zhu2024knowagent}. A representative method is RoG~\cite{luo2023reasoning}, which fine-tunes an LLM to generate reasoning relation paths and then retrieves the corresponding paths from the graph. 

While these retrievers have demonstrated success across different tasks, they remain unaware of the generation process, which often leads to incomplete retrieval and the introduction of substantial noise as shown in Section~\ref{sec:pre}. Moreover, training-based methods are usually tailored to a specific graph, further limiting their ability to generalize to unseen graphs. Instead, the proposed \method unifies retrieval and generation within the same process, enabling the model to actively explore the graph while constructing reasoning paths and generating answers in an integrated manner.

\subsection{Reasoning of Large Language Models}
Large Language Models (LLMs) have demonstrated strong performance on natural language tasks~\cite{chang2024survey, thirunavukarasu2023large, kasneci2023chatgpt, han2025reasoning}, but they may struggle with more complex reasoning problems~\cite{zheng2023large, nezhurina2024alice}. 
To further improve their reasoning ability, early works leverage prompting-based methods, such as Chain-of-Thought ~\cite{wei2022chain}, Tree-of-Thought~\cite{yao2023tree} and Graph-of-Thought ~\cite{besta2024graph}, which encourage models to generate and organize intermediate reasoning steps during inference. Think-on-graph (ToG)~\cite{sun2023think} also follow this paradigm to prompt LLM reason on the graph.
While effective, these methods are often sensitive to prompt design and computationally inefficient. Building on this, more recent approaches~\cite{xu2025towards, sui2025stop} focus on supervised fine-tuning (SFT) with curated reasoning traces~\cite{deng2021reasonbert, lobo2024impact} and reinforcement learning (RL) with process or outcome rewards~\cite{rafailov2023direct, zhang2024rest}. These methods allow models to internalize reasoning strategies rather than relying only on inference-time prompting. However, most of these advances focus on text-only tasks, whereas reasoning over graph-structured knowledge introduces additional challenges, requiring models to explore nodes, edges, and multi-hop paths. Although several works~\cite{wang2023keqing, luo2025graph, yu2025graphrag} have attempted to adapt LLMs for graph reasoning tasks, most of them explicitly decompose complex queries into simpler sub-queries and rely on external retrievers. It remains underexplored how to enable LLMs to automatically explore graphs and generate answers in a unified manner. 

In this paper, we follow previous studies~\cite{luo2023reasoning, he2024g, sun2023think} and leverage the Knowledge Graph Question Answering (KGQA) task to evaluate the effectiveness of graph reasoning, as it inherently involves multi-hop relational reasoning and explicit interaction with graph structures to obtain correct answers.

\section{Preliminary}
\label{sec:pre}
In this section, we first provide definitions, and 
then analyze the retrievers adopted in current popular GraphRAG methods.
\subsection{Definitions}
\textbf{Knowledge Graphs (KGs).} 
Let $\mathcal{G} = (\mathcal{V}, \mathcal{R}, \mathcal{E})$ be a knowledge graph, where $\mathcal{V}$ denotes entities, $\mathcal{R}$ relation types, and $\mathcal{E}$ directed edges. Each edge is represented as a triple $(h, r, t)$ with $h, t \in \mathcal{V}$ and $r \in \mathcal{R}$. For any node $v \in \mathcal{V}$, we denote its neighboring entities, incident relations, and connected edges as $\mathcal{N}_v$, $\mathcal{R}_v$, and $\mathcal{E}_v$, respectively.
In this paper, we also augment the knowledge graph with inverse edges. Specifically, for each triple $(h, r, t) \in \mathcal{E}$, we add its inverse edge $(t, r_{\text{inverse}}, h)$, following prior works~\cite{kazemi2018simple, dutt2022perkgqa}. 

\noindent \textbf{Knowledge Graph Question Answering (KGQA).} KGQA is the task of answering natural language questions by reasoning over a knowledge graph. Given a question and a KG, $\mathcal{G} = \{\mathcal{V}, \mathcal{R}, \mathcal{E}\}$, the goal is to identify the correct entity or set of entities in $\mathcal{V}$ that answer the query. This typically requires multi-hop reasoning, where the model must traverse nodes and relations along valid paths (e.g., $(h, r_1, v_1), (v_1, r_2, t)$) to reach the answer.

Due to the large size of knowledge graphs, which cannot fit into the limited context window of LLMs at once, most KGQA methods (i.e., GraphRAG) first retrieve a subgraph relevant to the query and then restrict the LLM’s reasoning to this subgraph. However, in this two-stage pipeline, the retriever and generator are usually optimized independently and executed sequentially. In the following subsections, we present experiments that illustrate the limitations of this design, focusing on the quality of the retrieved subgraphs, which can greatly affect the overall generation performance.
\subsection{Experimental Settings}
To conduct a comprehensive evaluation, we select one representative method from each retriever category as introduced in Section~\ref{sec:related}: G-Retriever~\cite{he2024g} for heuristic-based methods, GNN-RAG~\cite{mavromatis2024gnn} for GNN-based methods, and RoG~\cite{luo2023reasoning} for LLM-based methods.

We select two widely used datasets: WebQSP~\cite{yih2016value}, which primarily contains questions requiring one- or two-hop reasoning, and CWQ~\cite{talmor2018web}, which consists of more complex questions that require multi-hop reasoning. 
For each question, there can be multiple correct answers. On average, the number of answers per test question is 10.20 for WebQSP and 1.89 for CWQ. For the heuristic-based method, i.e., G-Retriever, no training of the retriever is required. In contrast, both GNN-RAG and RoG require training. Specifically, for GNN-RAG, we train a GNN model as the retriever on each dataset, while for RoG we fine-tune the Llama-3.1-8B-Instruct model~\cite{touvron2023llama} to serve as the retriever on each dataset. 
To evaluate retriever performance, we adopt three metrics: \textbf{Hit}, which measures whether at least one correct answer is included in the retrieved content; \textbf{Recall}, which measures the proportion of ground-truth answers retrieved for each question; and \textbf{Precision}, which measures the ratio of correctly retrieved entities among all retrieved entities.

\begin{figure*}[htb]
    \centering
    \begin{subfigure}[t]{0.49\textwidth}
        \centering
        \includegraphics[width=\linewidth]{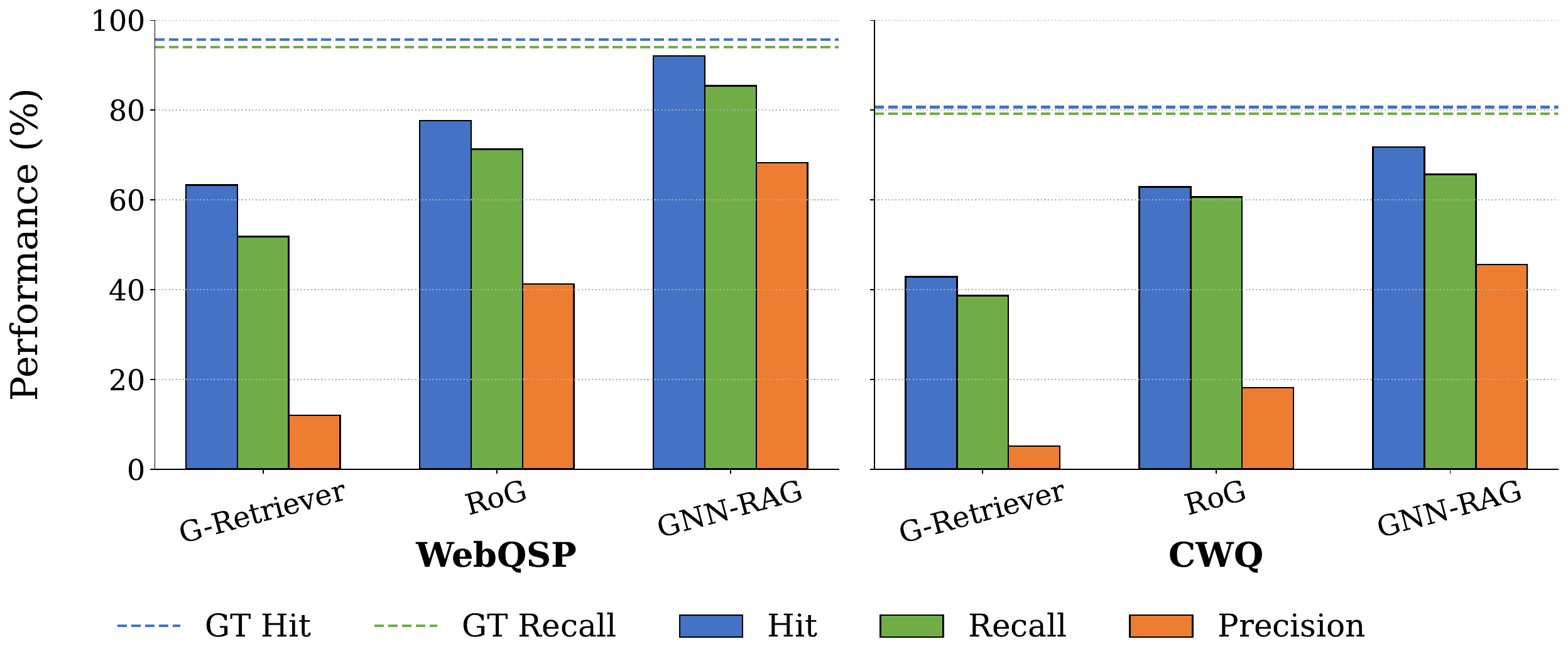}
        \caption{In-distribution performance comparison of three retrievers  on WebQSP and CWQ datasets. 
        Bars show Hit, Recall, and Precision, while dashed lines indicate ground-truth (GT) upper bounds. 
        }
        \label{fig:in_domain_results}
    \end{subfigure}
    \hfill
    \begin{subfigure}[t]{0.49\textwidth}
        \centering
        \includegraphics[width=\linewidth]{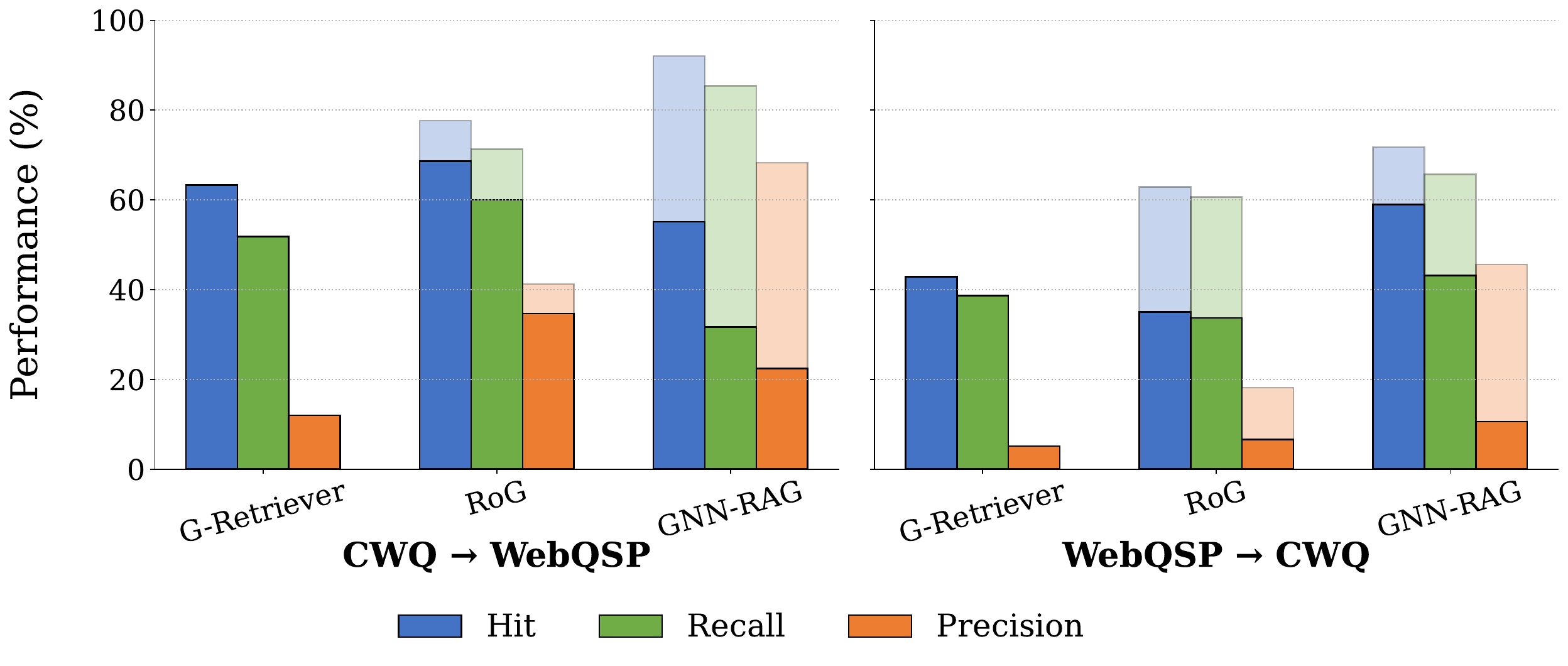}
        \caption{Cross-distribution generalization performance between CWQ and WebQSP. 
        Solid bars denote transfer results, while transparent bars show in-distribution results. 
        }
        \label{fig:cross_domain_results}
    \end{subfigure}
    \vspace{-0.1in}
    \caption{Retrieval and generalization performance of different methods on WebQSP and CWQ datasets.}
    \label{fig:retriever_overall}
    \vspace{-0.1in}
\end{figure*}
 \vspace{-0.1in}
\subsection{In-Distribution Retriever Performance}
\label{subsec:indisribution}
In this subsection, we aim to measure the performance of different retrievers when they are both trained and tested on the same dataset (i.e., WebQSP or CWQ). The performance of the three selected methods is shown in Figure~\ref{fig:in_domain_results}, where we also include the ground-truth Hit and Recall as upper bounds, since the underlying graph does not contain all the answers for every question.
From the results, we observe that none of the retrievers achieve perfect performance. For example, the Recall and Precision of G-Retriever on the CWQ dataset are only 38.65 and 5.14, respectively, indicating that many correct answers are missing while substantial noise is included in the retrieved subgraphs. 
Among the three methods, GNN-RAG performs best, but it still shows around a 10\% gap compared to the ground-truth Hit upper bound on the CWQ dataset. 
This demonstrates that the {\it limitations of current retrievers will substantially constrain downstream generation performance}.
\vspace{-0.1in}
\subsection{Cross-Distribution Generalization}
\label{subsec:crossdisrtibution}
In Section~\ref{subsec:indisribution}, we mainly studied the retriever performance when training and testing on the same dataset. However, in real-world applications it is often impractical to train a separate retriever for every new graph. In this subsection, we therefore explore the generalization ability of pretrained retrievers. Instead of fine-tuning the retrievers for each dataset as in GNN-RAG~\cite{mavromatis2024gnn} and RoG~\cite{luo2023reasoning}, we directly apply a retriever pretrained on one dataset to test on the other, in order to evaluate its cross-distribution generalization.



We first conduct experiments on the WebQSP and CWQ datasets, which are both built on Freebase KG~\cite{bollacker2008freebase} but differ substantially in question distribution.
WebQSP consists of short, naturally collected questions that require 1-2-hop reasoning,
whereas CWQ contains synthetically expanded questions with up to 4 hops and more complex linguistic structures such as conjunctions and comparatives.
As shown in Figure~\ref{fig:cross_domain_results}, both RoG and GNN-RAG suffer a significant performance drop under the cross-distribution setting.
For instance, the {Hit} of GNN-RAG decreases from 92.07 to 55.09, and its {Recall} drops from 85.43 to 31.68 on WebQSP when using the retriever trained on CWQ, performing even worse than the heuristic-based {G-Retriever}, which requires no retriever pretraining.

We further evaluate cross-graph generalization by applying retrievers trained on WebQSP to datasets with different knowledge graphs, including MetaQA~\cite{zhang2017variational}, which uses the WikiMovies KG, and PathQuestion~\cite{zhou2018interpretable}, which uses a Freebase subset with altered relation names. 
As shown in Table~\ref{tab:generalization_pre}, RoG fails to retrieve any valid content, 
while GNN-RAG retrieves overly large subgraphs with very low precision.
\vspace{-0.1in}
\begin{table}[H]
\centering
\caption{Cross-graph generalization performance.}
\label{tab:generalization_pre}
\vspace{-0.1in}
\resizebox{0.9\columnwidth}{!}{
\begin{tabular}{lccc|ccc}
\hline
 & \multicolumn{3}{c|}{Webqsp-\textgreater{}MetaQA} & \multicolumn{3}{c}{Webqsp-\textgreater{}PathQuestion} \\ \hline
 & Hit & Recall & Precision & Hit & Recall & Precision \\ \hline
\multicolumn{1}{c}{RoG} & 0 & 0 & 0 & 0 & 0 & 0 \\
\multicolumn{1}{c}{GNN-RAG} & 38.3 & 13.36 & 0.02 & 63.8 & 51.7 & 0.05 \\ \hline
\end{tabular}
}
\vspace{-0.1in}
\end{table}

Overall, these results show that representative GraphRAG retrievers generalize poorly.
We attribute this to two major factors: 
\begin{enumerate}[leftmargin=1.2em]
    \item \textbf{Distribution gap in questions.} 
    Retrievers are typically trained to recognize the lexical and structural patterns of questions within a single dataset (e.g., WebQSP or CWQ). 
    When the question style or reasoning composition changes, such as shifting from short, 1-2-hop natural questions to complex, synthetic multi-hop questions, the retriever fails to capture the correct reasoning paths and relevant entities.  This suggests that retrievers overfit to dataset-specific surface forms rather than learning transferable reasoning patterns.
    \item \textbf{Distribution gap in graphs.} 
    Differences in the underlying knowledge graphs, such as relation vocabulary, and entity type distributions, further limit generalization. 
    GNN-based retrievers depend on node and edge embeddings learned from a specific graph topology, 
    while LLM-based retrievers (e.g., RoG) rely on relation paths that may not align across graphs. 
    As a result, both classes of retrievers struggle to adapt when transferred to new graphs with different relation semantics or structural patterns.
\end{enumerate}

\vspace{-0.1in}
\subsection{Discussions}
\label{subsec:analysis}
 
From Section~\ref{subsec:indisribution}, we observe that a substantial portion of answers cannot be retrieved by existing retrievers.
Consequently, even strong generators fail to produce these correct answers since the necessary evidence is missing from the retrieved subgraphs. This highlights a key limitation of current two-stage GraphRAG paradigms—\textit{the retriever and generator operate independently and cannot interact with each other on the graph during reasoning}.

The results in Section~\ref{subsec:crossdisrtibution} further suggest that existing retrievers are over-specialized to their training datasets and knowledge graphs. They lack the capability to adjust their retrieval behavior based on the evolving environment or intermediate reasoning outcomes. \textit{This static retrieval behavior prevents effective adaptation to unseen question types and graph structures.}

These insights motivates a new perspective on GraphRAG:
rather than decoupling retrieval and generation or retrieving all potentially relevant content at once,
the model should be able to \textbf{interact with the whole graph step by step}, selecting and exploring new nodes and relations as reasoning progresses for the question.
Such a reasoning-driven retrieval process enables the model to dynamically incorporate feedback from intermediate steps,
allowing it to generalize across diverse questions and graph structures by learning to \textit{reason and explore}
instead of memorizing fixed retrieval patterns.


\section{Method}
\label{sec:method}
\begin{figure*}[!htb]
    \centering
    \includegraphics[width=0.95\linewidth, trim=20 10 20 10, clip]{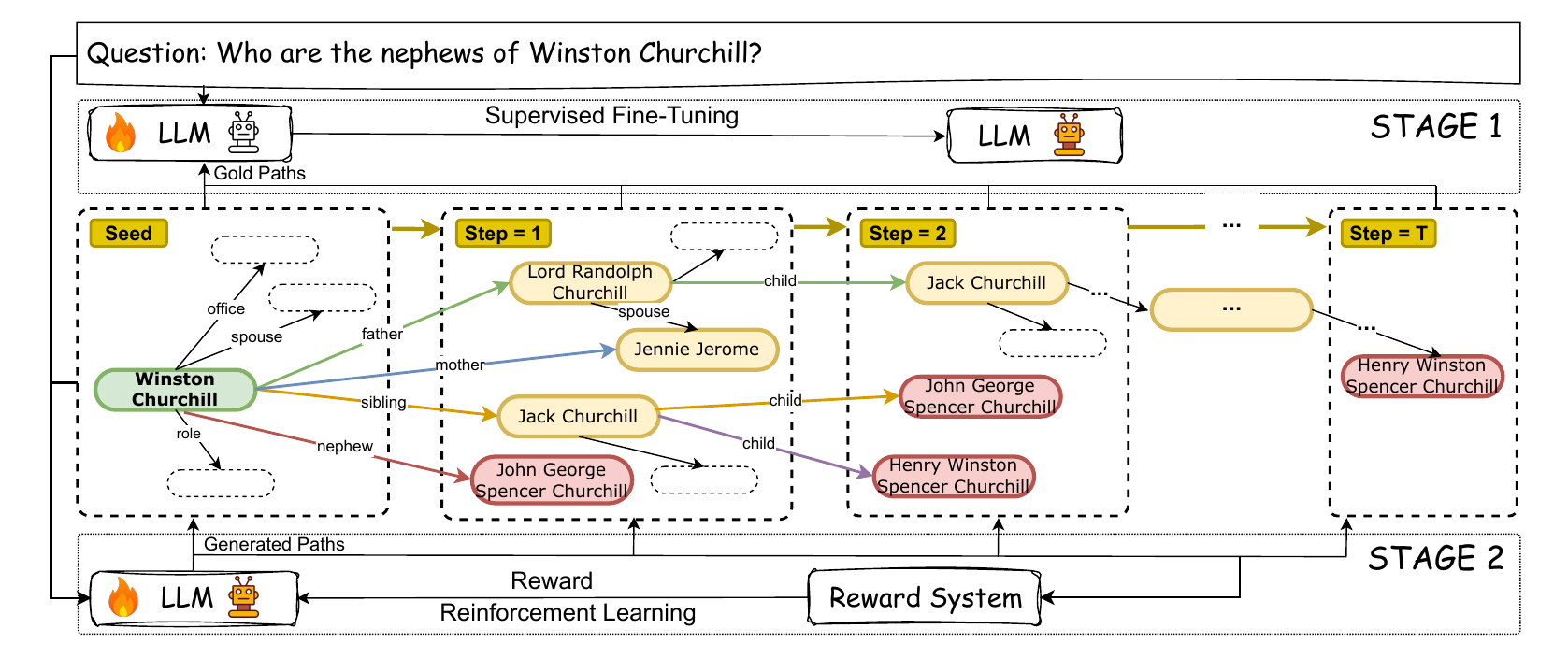}
    \vspace{-0.1in}
    \caption{The framework of \method. The model begins from the seed entity (green), incrementally expands to explored entities (yellow), while discovering answer entities (red). Stage 1 uses SFT to learn gold reasoning paths, and the trained model is then used as the initial model for Stage 2, where RL refines the exploration policy based on reward feedback.}
    \label{fig:roe}
    \vspace{-0.1in}
\end{figure*}

Motivated by the limitations of existing GraphRAG methods identified in Section~\ref{sec:pre}, we propose Reasoning by Exploration (\method), a unified framework that integrates retrieval and generation within a single process, as illustrated in Figure~\ref{fig:roe}. Unlike traditional two-stage GraphRAG pipelines, where a retriever extracts a static subgraph before reasoning begins, \method\ enables a single LLM to actively explore the entire knowledge graph step by step.
\vspace{-0.1in}
\subsection{Exploration Formulation}
\label{subsec:explore}
We formulate reasoning in \method\ as an \textbf{exploration process} over a knowledge graph, where the LLM incrementally traverses the graph to construct reasoning paths that lead to correct answers. 
Formally, we formulate the exploration process as follows: 

Given a question $q$ and a knowledge graph $\mathcal{G} = \{\mathcal{V}, \mathcal{R}, \mathcal{E}\}$, the model begins from one or more \textbf{seed entities} $\mathcal{V}_0$ that are linked to entities mentioned in the question. We expect the model to explore one-hop neighbors from the currently explored nodes at each step, progressively expanding the reasoning space. Specifically, at each step $d$, the model maintains a set of \textbf{current reasoning paths:} 
\[
\mathcal{P}_d = \{p_1^d, p_2^d, \ldots, p_n^d\},
\]
where each path $p_i^d$ is defined as
$p_i^d = [v_0, r_1, v_1, \ldots, r_k, v_k],$
starting from a seed entity $v_0 \in \mathcal{V}_0$ and expanding through a sequence of relations and entities. 
We define \textbf{frontier nodes} as
\[
\mathcal{F}_d = \{v_k \mid p_i^d \in \mathcal{P}_d\},
\]
representing the boundary entities that can be expanded at the next step. At step $d$, the model observes a \textbf{state}
\[
s_d = (q, \mathcal{P}_d, \mathcal{N}_{\mathcal{F}_d}),
\]
where $\mathcal{N}_{\mathcal{F}_d}$ denotes neighbor nodes and edges of all frontier nodes. 
The model then predicts an action $a$ that consists of two lists:
\begin{itemize}[leftmargin=1em]
    \item \textbf{Answers:} $\mathcal{A}_d = [ans_1^d, ans_2^d, \ldots, ans_m^d]$, representing entities predicted as current answers at step $d$.
    \item \textbf{New Exploration Paths:} $\mathcal{P}_{d+1} = \{p_i^d \oplus (v_k, r_{k+1}, v_{k+1}) \mid p_i^d = [v_0, r_1, v_1, \ldots, r_k, v_k] \in \mathcal{P}_d, (v_k, r_{k+1}, v_{k+1}) \in \mathcal{E}\}$, where $\oplus$ denotes the extension of an existing path with a newly explored triplet.
\end{itemize}
The exploration continues until the model predicts that no new paths should be expanded (i.e., $\mathcal{P}_{d+1} = \emptyset$) or a predefined maximum number of steps $D_{\max}$ is reached. 
The final outputs include all predicted answers
$\mathcal{A} = \bigcup_d \mathcal{A}_d$,
which enables the model to find multiple answers through reasoning paths of different lengths.

Through this design, retrieval and generation become mutually informed and jointly optimized, enabling LLMs to construct reasoning paths that are both complete and contextually relevant.
However, empowering LLMs to perform such structured graph exploration introduces several challenges. First, LLMs are pretrained primarily on unstructured text and therefore lack the inherent ability to navigate and reason over graphs~\cite{dai2024large}. Second, naively training an LLM on exploration trajectories may cause it to memorize specific graph patterns rather than learning transferable exploration strategies. To address these challenges, we introduce a two-stage training strategy that equips the model with both graph exploration ability and strong generalization across different graphs.
\subsection{Stage I: Supervised Fine-Tuning for Step-wise Exploration}
In the first stage, we aim to equip the LLM with the basic ability to perform step-wise graph exploration. Specifically, we require the \textbf{gold actions} for each step, which include both the step-wise answers and the new exploration paths as defined in Section~\ref{subsec:explore}. Previous works~\cite{zhang2022subgraph, sun2019pullnet} typically use the \textit{shortest path} from the query entities to the answers to supervise their models. However, we argue that relying solely on the shortest path is overly restrictive. In real knowledge graphs, there may exist multiple valid reasoning paths connecting the same query entity to the correct answer; constraining training to only the shortest one discourages exploration and limits generalization. 
Moreover, the model does not have a global view of the entire graph at each step. When several neighboring entities share the same relation with the current frontier entity, selecting only a single target node can be ambiguous, especially when entity names are represented as IDs.

To address these issues, we construct the gold exploration paths by including all reasoning paths whose lengths are below a threshold, ensuring that the model learns to explore multiple plausible reasoning routes. Additionally, when one node is selected for expansion, other nodes connected to the frontier entity through the same relation are also treated as valid expansion candidates.
For the answers at each step, we use the newly explored entities that correspond to the ground-truth answers as the gold answers. The detailed procedure is summarized in Algorithm~\ref{alg:mine_and_sft} in Appendix~\ref{sec:appendix:a}.

In the supervised fine-tuning (SFT) stage, each training instance is represented as a pair $(s_d, a_d^{*})$ in $\mathcal{D}_{\text{SFT}}$, where $s_d$ denotes the state at step $d$ (including the question, current reasoning paths, and local neighborhood), 
and $a_d^{*}$ is the corresponding gold action constructed in Algorithm~\ref{alg:mine_and_sft}. 
At each step, the model takes $s_d$ as input and predicts action $a_d = (\mathcal{A}_d, \mathcal{P}_d)$, 
which includes the step-wise answers and the new exploration paths. 
The training objective is to maximize the likelihood of the gold action $a_d^{*}$ given its state, leading to loss:
\[
\mathcal{L}_{\text{SFT}} = -\!\sum_d \log P(a_d^{*} \mid s_d;\theta),
\]
where $\theta$ denotes the model parameters. 
This loss encourages the model to imitate the expert exploration behavior, 
thereby acquiring the fundamental ability for structured 
graph exploration before reinforcement-learning refinement in Stage~II.

\subsection{Stage II: Reinforcement Learning for Exploration Optimization}

Although the SFT stage enables the model to imitate gold trajectories, it may cause the LLM to memorize fixed exploration patterns within a single dataset, limiting its ability to generalize to unseen graphs or question types. To address this, we introduce a reinforcement learning (RL) stage that allows the model to interact with the knowledge graph and refine its exploration strategy through feedback rather than imitation.
During this stage, we not only encourage the LLM to generate correct step-wise answers and exploration paths, but also aim to promote broader and more diverse exploration behaviors, enabling the model to discover additional valid reasoning trajectories and answers beyond those seen in the gold data. To enable such flexible and interpretable optimization, we adopt a rule-based reinforcement learning framework, which provides explicit feedback signals without relying on a learned value model.
This choice is inspired by recent advances in reasoning LLMs, such as DeepSeek-R1~\cite{guo2025deepseek}, which have shown that rule-based rewards can effectively guide reasoning and exploration, while simplifying training compared to the original RLHF framework~\cite{ouyang2022training}. 

In \method, we expect the model to generate outputs in the correct format, align its predictions with the gold answers and exploration paths, and simultaneously discover new valid answers and paths while penalizing hallucinated ones. Therefore, we define the following five rewards:


\noindent \textbf{Format Reward.}  
Different from previous methods that rely on a ``thinking'' process, 
we directly prompt the LLM to generate both the step-wise answers and exploration paths 
in a structured \texttt{JSON} format to improve efficiency. Specifically, at each step the model is required to output:
\[
\{\texttt{"answers": [\,], "exploration\_paths": [\,]}\}.
\]
A non-zero reward $\mathcal{R}_\text{format}$ is given only when the model successfully follows the required format. 

\noindent \textbf{Answer Reward.}  
We measure how many of the gold answers $\mathcal{A}_d^{*}$ in the current state $s_d$ are successfully predicted by the model.  
Specifically, we compute the recall between the predicted and gold answer sets at each step $d$:
\[
\mathcal{R}_{\text{ans}} = 
\frac{|\mathcal{A}_d^p \cap \mathcal{A}_d^{*}|}{|\mathcal{A}_d^{*}|},
\]
where $\mathcal{A}_d^p$ denotes the predicted answers.  
A higher reward is assigned when more ground-truth answers are correctly identified.

\noindent \textbf{Answer Discovery Reward.}
To promote broader and more effective exploration, we further encourage the model to discover new valid answers 
that were not present in the previous steps.  
Let $\mathcal{A}$ denote the set of all gold answers for the question.  
We reward the model for predicting new correct answers 
$\mathcal{A}_d^p \cap (\mathcal{A} \setminus \mathcal{A}_d^{*})$ 
and penalize it for generating invalid ones $\mathcal{A}_d^p \setminus \mathcal{A}$:
\[
\mathcal{R}_{\text{ans-dis}} = 
|\mathcal{A}_d^p \cap (\mathcal{A} \setminus \mathcal{A}_d^{*})| 
- \beta \cdot |\mathcal{A}_d^p \setminus \mathcal{A}|,
\]
where $\beta$ is a penalty weight controlling the impact of hallucinated predictions.  This reward encourages the model to find novel correct answers, by leveraging its internal knowledge and the reasoning paths explored so far, while maintaining answer precision.

\noindent \textbf{Exploration Reward.}  
Similar to the answer reward, this reward evaluates how well the model’s predicted reasoning paths match the gold paths in the current state.  
Let $\mathcal{P}_d^{p}$ and $\mathcal{P}_d^{*}$ denote the predicted and gold reasoning paths at step $d$, respectively. 
We compute the reward as the recall between the predicted and gold path sets:
\[
\mathcal{R}_{\text{explore}} = 
\frac{|\mathcal{P}_d^{p} \cap \mathcal{P}_d^{*}|}{|\mathcal{P}_d^{*}|}.
\]
A higher reward indicates that the model successfully explores the correct reasoning paths in the current step, 
guiding it to perform faithful and consistent graph traversal.

\noindent \textbf{Exploration Discovery Reward.}
To further promote effective and diverse reasoning, we encourage the model to explore new valid paths while penalizing invalid ones.  
Let $\mathcal{P}_d^{p}$ denote the set of predicted paths at step $d$, and $\mathcal{P}_d^{*}$ denote the set of gold paths.  
We define \textit{new paths} as those that do not appear in $\mathcal{P}_d^{*}$ but exist on the knowledge graph $\mathcal{G}$, 
and \textit{invalid paths} as those whose triplets are not present in $\mathcal{G}$.  
The reward is computed as:
\[
\mathcal{R}_{\text{exp-dis}} = 
|\mathcal{P}_d^{p} \setminus \mathcal{P}_d^{*}|_{\text{valid}} 
- \beta \cdot |\mathcal{P}_d^{p}|_{\text{invalid}},
\]
where $|\mathcal{P}_d^{p} \setminus \mathcal{P}_d^{*}|_{\text{valid}}$ counts the number of new but valid paths on the graph 
and $|\mathcal{P}_d^{p}|_{\text{invalid}}$ counts the number of paths containing nonexistent triplets.  
This reward encourages the model to explore novel and meaningful reasoning directions while discouraging invalid graph traversals.

\noindent \textbf{Final Reward.}  
The overall reward for each step is defined as the sum of all the individual rewards introduced above:
\[
\mathcal{R}_{\text{total}} 
= \mathcal{R}_{\text{format}} 
+ \mathcal{R}_{\text{ans}} 
+ \mathcal{R}_{\text{ans-dis}} 
+ \mathcal{R}_{\text{explore}} 
+ \mathcal{R}_{\text{exp-dis}}.
\]
This combined reward jointly encourages the model to produce well-structured outputs, 
generate correct answers, and discover diverse and valid reasoning paths during reinforcement learning. We optimize \method\ using {Group Relative Policy Optimization (GRPO)}~\cite{guo2025deepseek}, 
a variant of Proximal Policy Optimization (PPO)~\cite{schulman2017proximal}. 


In summary, \method\ is first supervised through step-wise fine-tuning (Stage~I) 
to learn the fundamental exploration behavior from gold trajectories, 
and is then refined via reinforcement learning (Stage~II) 
to improve its reasoning and generalization ability.  
Through this two-stage training framework, 
\method\ learns to jointly reason and explore over graphs, 
constructing accurate and diverse reasoning paths while producing 
faithful multi-hop answers.


\section{Experiments}
\label{sec:exp}
In this section, we conduct comprehensive experiments to validate the effectiveness of the proposed \method. Specifically, we aim to answer the following research questions:
\textbf{(1) RQ1:} How does \method perform on benchmark datasets compared with baseline methods? \textbf{(2) RQ2:} Can \method generalize to unseen datasets better than existing GraphRAG approaches?  \textbf{(3) RQ3:} How do different design choices and factors influence the performance of \method?

\begin{table*}[!ht]
\centering
\caption{Performance comparison on two KGQA benchmarks. The best and second-best are in \textbf{bold} and \underline{underline}. \\ $^*$ indicates the model uses Llama3.1-8B-Instruct as backbone.}
\vspace{-0.05in}
\label{tab:overall}
\begingroup
\setlength{\tabcolsep}{6pt}        
\renewcommand{\arraystretch}{0.9} 

\begin{tabular}{c|c|cc|cc|c}
\hline
\multirow{2}{*}{\textbf{Category}} &
\multirow{2}{*}{\textbf{Method}}   &
\multicolumn{2}{c|}{\textbf{WebQSP}} &
\multicolumn{2}{c|}{\textbf{CWQ}}    &
\multirow{2}{*}{\textbf{Avg. Rank}} \\ 
& & \textbf{Hits@1} & \textbf{F1} & \textbf{Hits@1} & \textbf{F1} & \\ 
\hline
\multirow[c]{5}{*}{LLMs-only}
 & Flan-T5-xl                 & 31.00 &   -    & 14.70 &   -    & 12.00 \\
 & Alpaca-7B                  & 51.80 &   -    & 27.40 &   -    & 11.00 \\
 & Llama-2-7B-Chat            & 64.40 &   -    & 34.60 &   -    & 8.50 \\
 & Llama-3.1-8B-Instruct$^*$  & 58.96 & 31.66  & 31.49 & 20.66  & 10.25 \\
 & GPT-4.1                    & 64.99 & 44.04  & 44.09 & 36.56  & 8.25 \\
\hdashline
\multirow[c]{3}{*}{GNNs-only}
 & GraftNet                   &   -   & 62.40  &   -   & 32.70  & 8.00 \\
 & SR+NSM                     &   -   & 64.10  &   -   & 47.10  & 5.50 \\
 & UniKGQA                    &   -   & 70.20  &   -   & 48.00  & 3.50 \\
\hdashline
\multirow[c]{2}{*}{LLMs+GNNs}
 & G-Retriver$^*$             & 74.07 & 54.51  & 51.51 & 45.18  & 6.75 \\
 & GNNRAG$^*$                 & 85.25 & \underline{72.01} & \underline{63.61} & \textbf{54.92} & \underline{2.00} \\
\hdashline
\multirow[c]{5}{*}{LLMs+KGs}
 & ToG + Llama2-70B-Chat      & 63.70 &   -    & 53.60 &   -    & 7.50 \\
 & RoG$^*$                    & 78.62 & 64.39  & 53.83 & 45.97  & 5.00 \\
 & KD-CoT$^*$                 & 68.60 & 52.50  & 55.70 &   -    & 6.33 \\
 & SubGraphRAG$^*$            & \underline{86.61} & 70.57 & 56.98 & 47.16 & 3.00 \\
 & \textbf{RoE}$^*$                    & \textbf{89.13} & \textbf{74.76} & \textbf{66.49} & \underline{53.21} & \textbf{1.25} \\
\hline
\end{tabular}
\endgroup
\end{table*}

\subsection{Experimental Settings}
\textbf{Datasets.} To evaluate the effectiveness of \method, we use two widely adopted datasets:  \textbf{WebQSP}~\cite{yih2016value} and \textbf{CWQ}~\cite{talmor2018web}, both constructed upon the Freebase knowledge graph~\cite{bollacker2008freebase}.  
To further assess the generalization ability, we additionally use \textbf{MetaQA}~\cite{zhang2017variational}, which is built on the WikiMovies KG~\cite{miller2016key}, 
and \textbf{PathQuestion}~\cite{zhou2018interpretable}, which uses a Freebase subset with modified relation names. 


\noindent \textbf{Evaluation Metrics.} 
We follow previous works~\cite{he2021improving, guo2025empowering, li2024simple} 
and adopt two evaluation metrics: \textbf{Hit} and \textbf{F1}.  Specifically, the Hit measures the proportion of questions for which at least one correct predicted answer, while F1 provides a balanced evaluation of both precision and recall.

\noindent \textbf{Baselines.}  
We compare \method\ with a diverse set of baselines, which can be grouped into four categories:
(1) \textbf{LLM-Only:} methods that rely solely on the internal knowledge of LLMs to answer questions without using external knowledge graphs. 
We include {Flan-T5-XL}~\cite{chung2024scaling}, {Alpaca-7B}~\cite{taori2023stanford}, {Llama-2-7B-Chat}~\cite{touvron2023llama}, {Llama-3.1-8B-Instruct}~\cite{touvron2023llama}, and {GPT-4.1}~\cite{achiam2023gpt}.  
(2) \textbf{GNN-Only:} models that use GNNs for both retrieval and reasoning. 
We include {GraftNet}~\cite{sun2018open}, {SR+NSM}~\cite{zhang2022subgraph}, and {UniKGQA}~\cite{jiang2022unikgqa}.   
(3) \textbf{LLMs + GNNs:} hybrid approaches that combine LLMs and GNNs for retrieval and generation. 
For instance, {G-Retriever}~\cite{he2024g} leverages GNNs to enhance LLM reasoning, 
while {GNN-RAG}~\cite{mavromatis2024gnn} employs GNNs for retrieval and LLMs for generation. (4) \textbf{LLMs + KG:} methods that employ LLMs for both retrieval and generation over knowledge graphs. 
We compare with {ToG}~\cite{sun2023think}, {RoG}~\cite{luo2023reasoning}, {KD-CoT}~\cite{wang2023knowledge}, and {SubGraphRAG}~\cite{li2024simple}. 

\noindent \textbf{\method settings.}  
For \method, we use the Llama-3.1-8B-Instruct model~\cite{touvron2023llama} as the backbone.  
We divide the training data of both WebQSP and CWQ into two parts: 60\% is used for the SFT (stage I) and the remaining 40\% for the RL (stage II). To reduce memory consumption during training, we apply the LoRA~\cite{hu2022lora} in both stages. 

To ensure a fair comparison, we train each retriever and generator on a single dataset, rather than jointly training across multiple datasets as done in RoG and GNN-RAG.  
Moreover, to eliminate the influence of backbone model, 
we use the same \textbf{Llama-3.1-8B-Instruct} model~\cite{touvron2023llama}
for most LLM-based baselines, including RoG, KD-CoT, SubGraphRAG, G-Retriever, and GNN-RAG. We also evaluate with different backbone models in Appendix~\ref{app:sec:backbone}.



\vspace{-0.1in}
\subsection{Overall Performance Comparison}
The overall performance on the WebQSP and CWQ datasets is shown in Table~\ref{tab:overall}. 
We make the following observations: 
\begin{itemize}[leftmargin=1.5em]
    \item The proposed \method\ achieves the best performance across most metrics on both datasets. 
    Specifically, \method\ attains relative improvements of 2.9\% and 3.8\% on Hit and F1, respectively, 
    compared to the second-best models on the WebQSP dataset. 
    On the CWQ dataset, \method\ also achieves a 3.1\% relative improvement in Hit,
    demonstrating its strong ability to reason on graphs.
    \item The LLM-only models, which lack access to graph information, generally perform worse than methods that incorporate graphs. However, stronger models, such as GPT-4.1, achieve better results, suggesting that large LLMs already possess internal knowledge relevant to the benchmarks.
    This observation justifies our choice that using the same LLM backbone for baselines to ensure a fair comparison.
    \item The GNN-only models generally perform worse than methods that incorporate LLMs and graph information, indicating that the internal knowledge and reasoning capability of LLMs are essential for solving multi-hop KGQA tasks.
    \item Compared with LLMs+GNNs methods, \method achieves better performance in most cases, indicating additional GNN modules may not be necessary, suggesting LLMs themselves can acquire effective graph reasoning capabilities through fine-tuning.
    \item For the LLMs+KGs methods, such as ToG, KD-CoT, and SubGraphRAG, which prompt pretrained LLMs to iteratively retrieve and generate sub-questions or directly answer based on a retrieved subgraph, the performance is still lower than that of the proposed \method. This suggests that pretrained LLMs are not well-suited for graph reasoning without additional fine-tuning.
\end{itemize}

\subsection{Generalization Performance Comparison}

\begin{figure*}[htb]
    \centering
    \begin{subfigure}[t]{0.32\textwidth}\centering
        \includegraphics[width=\linewidth]{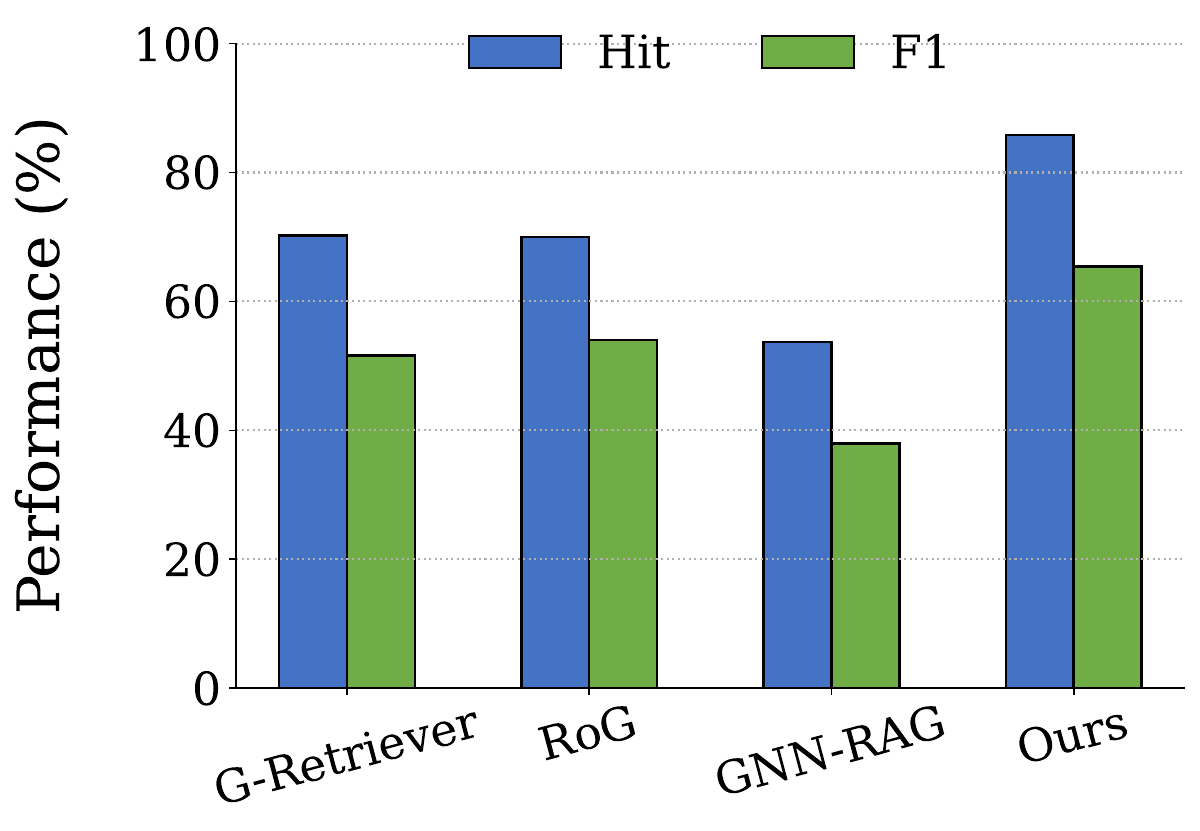}
        \caption{CWQ → WebQSP}
    \end{subfigure}\hfill
    \begin{subfigure}[t]{0.32\textwidth}\centering
        \includegraphics[width=\linewidth]{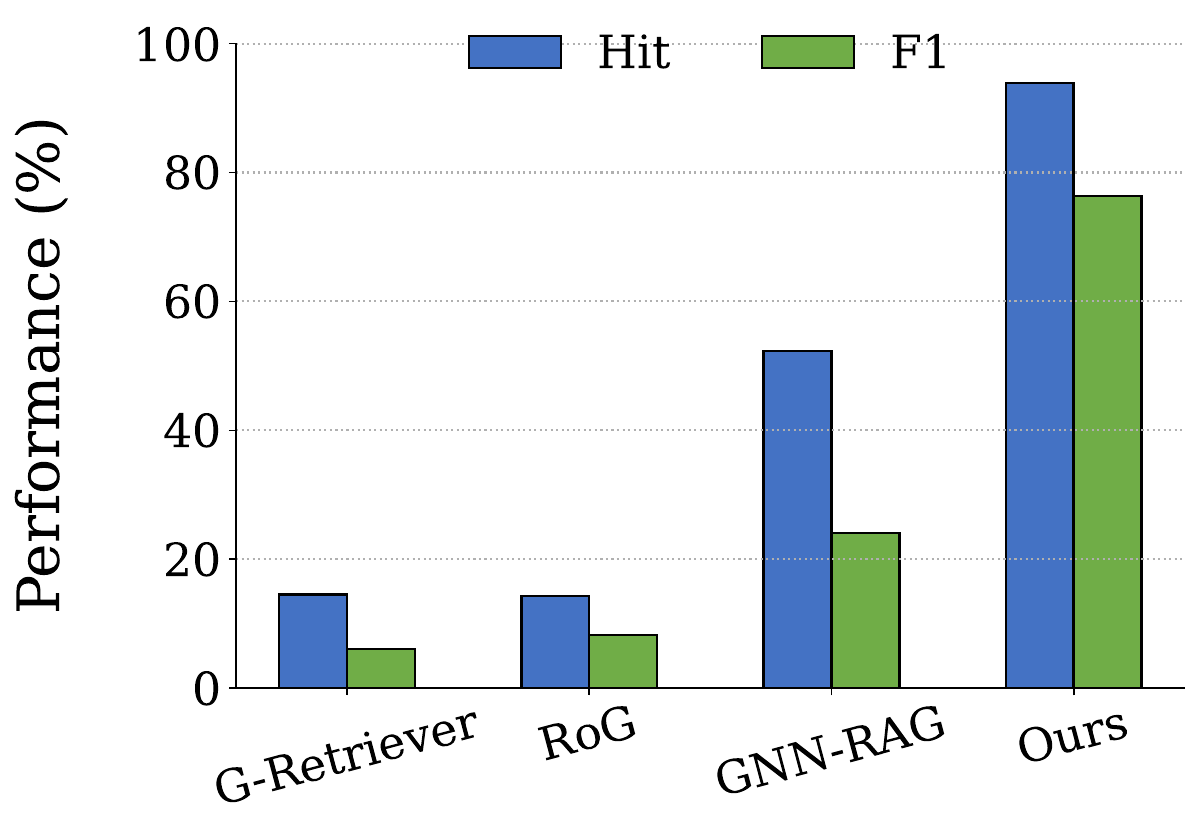}
        \caption{WebQSP → MetaQA}
    \end{subfigure}\hfill
    \begin{subfigure}[t]{0.32\textwidth}\centering
        \includegraphics[width=\linewidth]{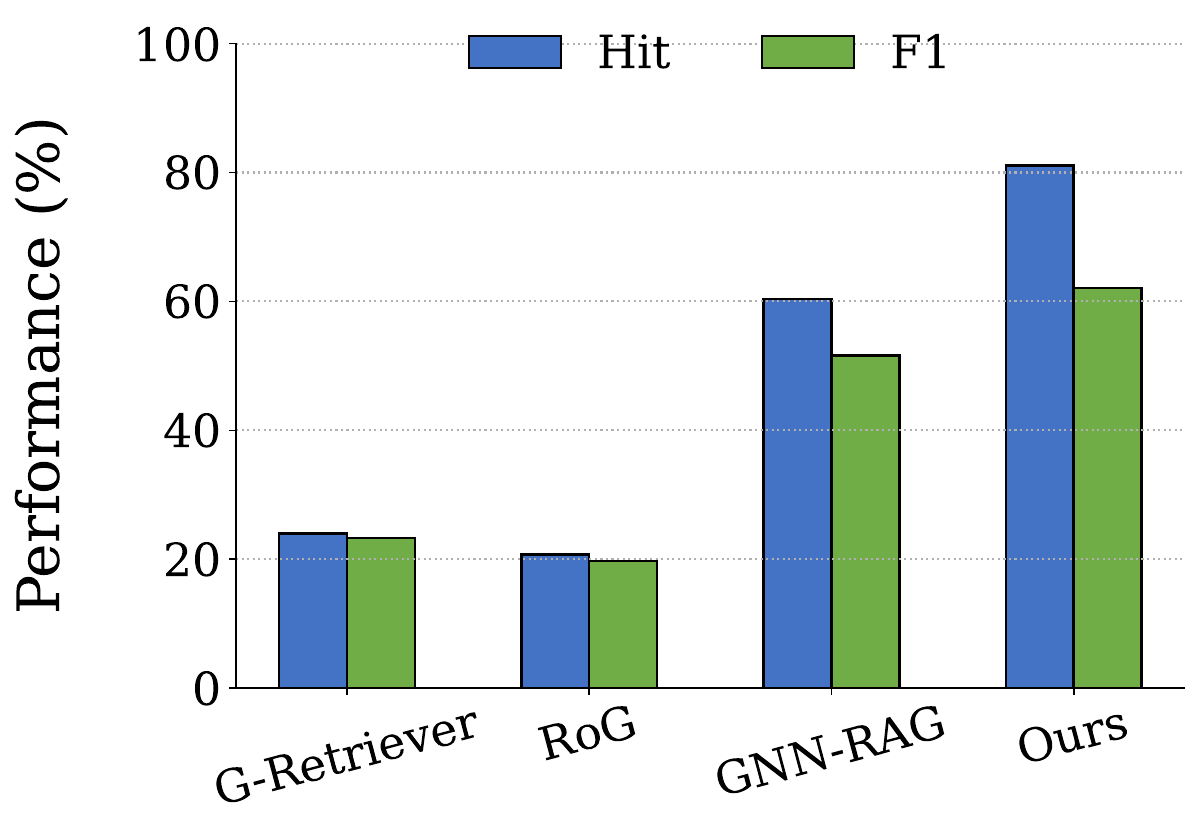}
        \caption{WebQSP → PathQuestion}
    \end{subfigure}
    \vspace{-0.1in}
    \caption{Generalization performance (Hit/F1) of different methods across dataset transfers.}
    \label{fig:retriever_generalization}
\end{figure*}

In this section, we evaluate the generalization capability of \method.  
Unlike previous works~\cite{luo2023reasoning, mavromatis2024gnn}, which fine-tune their models on the target graphs, we directly apply the pretrained model from one dataset to unseen datasets without any additional training.  
Specifically, we select {G-Retriever}, {RoG}, and {GNN-RAG} as baselines, and evaluate the transfer performance under three settings: \textbf{CWQ}~$\rightarrow$~\textbf{WebQSP}, \textbf{WebQSP}~$\rightarrow$~\textbf{PathQuestion}, and \textbf{WebQSP}~$\rightarrow$~\textbf{MetaQA}.

The generalization results are presented in Figure~\ref{fig:retriever_generalization}. As shown, the proposed \method\ significantly outperforms all baselines across different transfer settings. For example, under the {WebQSP~$\rightarrow$~MetaQA} transfer, \method\ achieves a {Hit} of \textbf{93.90}, while the second-best method only reaches \textbf{52.30}, demonstrating the strong generalization capability of \method\ across distributional shifts in both questions and knowledge graphs. Specifically, although {G-Retriever} employs a heuristic-based retriever without explicit retriever training, it still requires training a GNN model to encode the graph structure.
As a result, G-Retriever performs well on the {CWQ~$\rightarrow$~WebQSP} transfer setting, where both datasets share the same underlying knowledge graph, but performs poorly in other settings with different graph structures.
Similarly, the {RoG} model relies on predicting reasoning relation paths, which are difficult to transfer across graphs with different relation vocabularies. {GNN-RAG} relies on trained GNNs for retrieval, which also makes it difficult to generalize to different question distributions and knowledge graphs.
As a result, it tends to retrieve overly large subgraphs under transfer settings, as discussed in Section~\ref{sec:pre}, thereby introducing substantial noise that negatively affects the generator’s reasoning performance.

\subsection{Ablation Studies}

In this subsection, we conduct ablation studies to verify the effectiveness of the key components in \method.  
Specifically, we analyze the impact of the reinforcement learning (RL) stage and the proposed discovery rewards for both answers and exploration paths. We conduct experiments on two variants of \method: (1) removing the reinforcement learning (RL) stage and using only supervised fine-tuning (SFT), denoted as \textbf{\method~w/o~RL}; and
(2) removing the discovery rewards for both Answer Discovery Reward and Exploration Discovery Reward during the RL stage, denoted as \textbf{\method~w/o~Discovery}. We train these variants on WebQSP dataset and test their performance on WebQSP, MetaQA and PathQuestion dataset. 
\begin{figure}[!htb]
    \centering
    \includegraphics[width=0.9\linewidth]{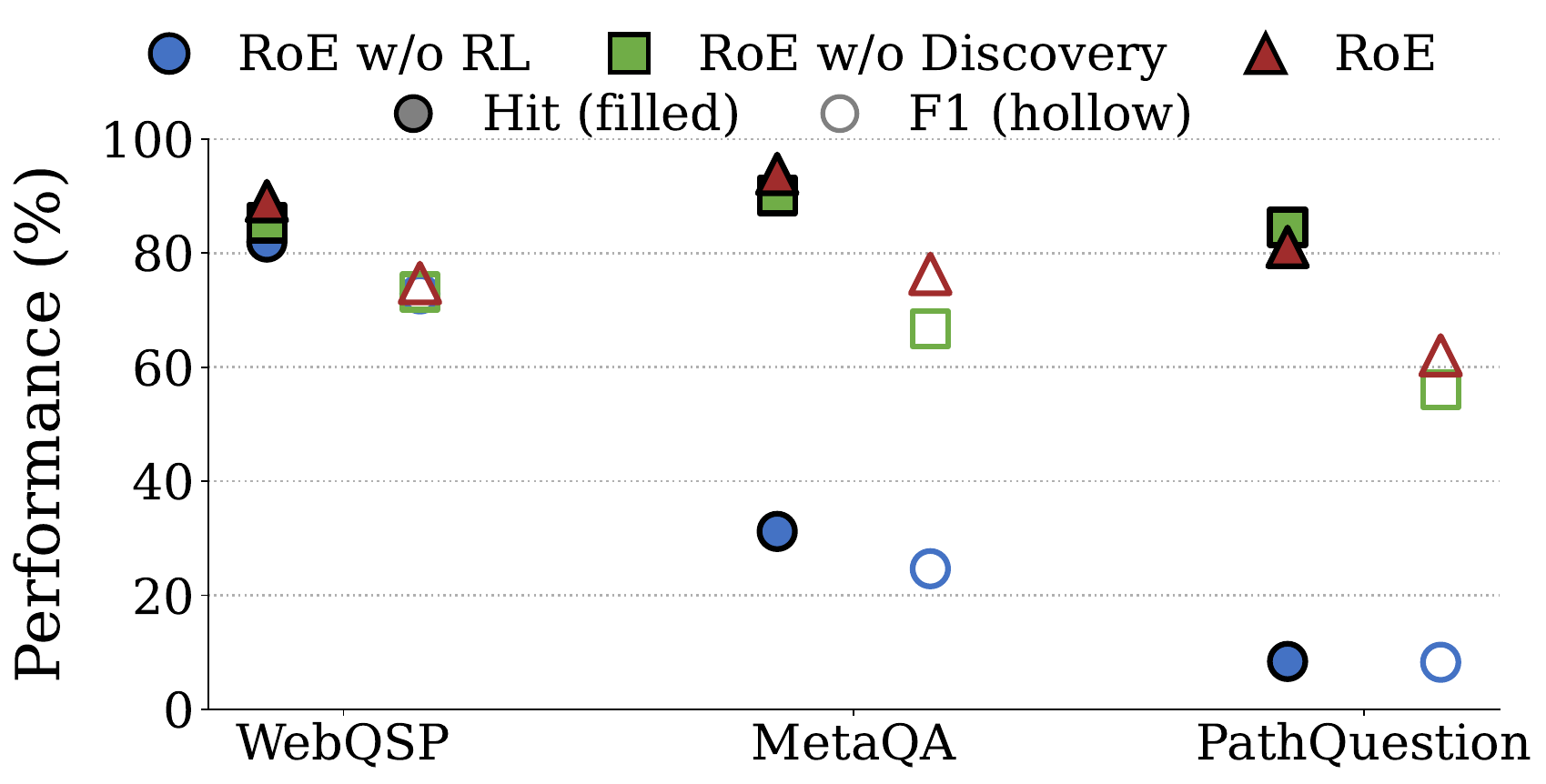}
    \vspace{-0.1in}
    \caption{Performance of \method and its variants pretrained on the WebQSP dataset.
Marker shapes denote different model variants, while filled and hollow markers represent Hit and F1 metrics, respectively.}
    \label{fig:ablation}
\vspace{-0.1in}
\end{figure}

The results are presented in Figure~\ref{fig:ablation}, from which we make the following observations:
\textbf{(1)} The RL stage is essential for \method, especially under transfer settings.
\method~w/o~RL shows significantly lower performance on the MetaQA and PathQuestion datasets, demonstrating that models trained only with SFT tend to memorize dataset-specific patterns and struggle to generalize to unseen graphs without reinforcement learning.
\textbf{(2)} The discovery rewards, which guide the model to find more valid answers and paths while penalizing hallucinations, are also crucial for performance.
Without these rewards, the model tends to generate more hallucinated answers, resulting in a notable drop in F1 under transfer settings.

\section{Conclusion}
In this paper, we revisited the limitations of existing two-stage GraphRAG frameworks, 
where the retriever and generator are optimized independently and lack interaction during reasoning.  
We proposed \textbf{Reasoning by Exploration (\method)}, a unified framework that integrates retrieval and generation into a step-wise exploration process.  
By allowing the model to interact with the graph dynamically and incorporate feedback from intermediate steps, 
\method\ learns to reason and explore rather than rely on static retrieval patterns.  
Extensive experiments on multiple KGQA benchmarks demonstrate that \method\ consistently outperforms strong baselines 
and generalizes effectively across different question distributions and graphs.

\section*{Acknowledgment}
This work is supported by the National Science Foundation (NSF) under grant numbers CNS2321416, IIS2212032, IIS2212144, IIS2504089, DUE2234015, CNS2246050, DRL2405483, IIS 2524379, and NAIRR 250188, the Michigan Department of Agriculture and Rural Development, US Dept of Commerce, Gates Foundation, Amazon Faculty Award, Meta, NVIDIA, Microsoft and SNAP.


\bibliographystyle{ACM-Reference-Format}
\balance
\bibliography{mybib}

@article{chang2024survey,
  title={A survey on evaluation of large language models},
  author={Chang, Yupeng and Wang, Xu and Wang, Jindong and Wu, Yuan and Yang, Linyi and Zhu, Kaijie and Chen, Hao and Yi, Xiaoyuan and Wang, Cunxiang and Wang, Yidong and others},
  journal={ACM transactions on intelligent systems and technology},
  volume={15},
  number={3},
  pages={1--45},
  year={2024},
  publisher={ACM New York, NY}
}

@inproceedings{wang2020traffic,
  title={Traffic flow prediction via spatial temporal graph neural network},
  author={Wang, Xiaoyang and Ma, Yao and Wang, Yiqi and Jin, Wei and Wang, Xin and Tang, Jiliang and Jia, Caiyan and Yu, Jian},
  booktitle={Proceedings of the web conference 2020},
  pages={1082--1092},
  year={2020}
}

@article{thirunavukarasu2023large,
  title={Large language models in medicine},
  author={Thirunavukarasu, Arun James and Ting, Darren Shu Jeng and Elangovan, Kabilan and Gutierrez, Laura and Tan, Ting Fang and Ting, Daniel Shu Wei},
  journal={Nature medicine},
  volume={29},
  number={8},
  pages={1930--1940},
  year={2023},
  publisher={Nature Publishing Group US New York}
}

@article{kasneci2023chatgpt,
  title={ChatGPT for good? On opportunities and challenges of large language models for education},
  author={Kasneci, Enkelejda and Se{\ss}ler, Kathrin and K{\"u}chemann, Stefan and Bannert, Maria and Dementieva, Daryna and Fischer, Frank and Gasser, Urs and Groh, Georg and G{\"u}nnemann, Stephan and H{\"u}llermeier, Eyke and others},
  journal={Learning and individual differences},
  volume={103},
  pages={102274},
  year={2023},
  publisher={Elsevier}
}

@article{lewis2020retrieval,
  title={Retrieval-augmented generation for knowledge-intensive nlp tasks},
  author={Lewis, Patrick and Perez, Ethan and Piktus, Aleksandra and Petroni, Fabio and Karpukhin, Vladimir and Goyal, Naman and K{\"u}ttler, Heinrich and Lewis, Mike and Yih, Wen-tau and Rockt{\"a}schel, Tim and others},
  journal={Advances in neural information processing systems},
  volume={33},
  pages={9459--9474},
  year={2020}
}

@article{gao2023retrieval,
  title={Retrieval-augmented generation for large language models: A survey},
  author={Gao, Yunfan and Xiong, Yun and Gao, Xinyu and Jia, Kangxiang and Pan, Jinliu and Bi, Yuxi and Dai, Yixin and Sun, Jiawei and Wang, Haofen and Wang, Haofen},
  journal={arXiv preprint arXiv:2312.10997},
  volume={2},
  number={1},
  year={2023}
}

@inproceedings{jiang2023active,
  title={Active retrieval augmented generation},
  author={Jiang, Zhengbao and Xu, Frank F and Gao, Luyu and Sun, Zhiqing and Liu, Qian and Dwivedi-Yu, Jane and Yang, Yiming and Callan, Jamie and Neubig, Graham},
  booktitle={Proceedings of the 2023 Conference on Empirical Methods in Natural Language Processing},
  pages={7969--7992},
  year={2023}
}

@article{xu2024ram,
  title={Ram-ehr: Retrieval augmentation meets clinical predictions on electronic health records},
  author={Xu, Ran and Shi, Wenqi and Yu, Yue and Zhuang, Yuchen and Jin, Bowen and Wang, May D and Ho, Joyce C and Yang, Carl},
  journal={arXiv preprint arXiv:2403.00815},
  year={2024}
}

@inproceedings{wiratunga2024cbr,
  title={CBR-RAG: case-based reasoning for retrieval augmented generation in LLMs for legal question answering},
  author={Wiratunga, Nirmalie and Abeyratne, Ramitha and Jayawardena, Lasal and Martin, Kyle and Massie, Stewart and Nkisi-Orji, Ikechukwu and Weerasinghe, Ruvan and Liret, Anne and Fleisch, Bruno},
  booktitle={International Conference on Case-Based Reasoning},
  pages={445--460},
  year={2024},
  organization={Springer}
}

@inproceedings{zhang2023enhancing,
  title={Enhancing financial sentiment analysis via retrieval augmented large language models},
  author={Zhang, Boyu and Yang, Hongyang and Zhou, Tianyu and Ali Babar, Muhammad and Liu, Xiao-Yang},
  booktitle={Proceedings of the fourth ACM international conference on AI in finance},
  pages={349--356},
  year={2023}
}

@article{chu2025enhancing,
  title={Enhancing LLM-Based Short Answer Grading with Retrieval-Augmented Generation},
  author={Chu, Yucheng and He, Peng and Li, Hang and Han, Haoyu and Yang, Kaiqi and Xue, Yu and Li, Tingting and Krajcik, Joseph and Tang, Jiliang},
  journal={arXiv preprint arXiv:2504.05276},
  year={2025}
}

@article{wu2020comprehensive,
  title={A comprehensive survey on graph neural networks},
  author={Wu, Zonghan and Pan, Shirui and Chen, Fengwen and Long, Guodong and Zhang, Chengqi and Philip, S Yu},
  journal={IEEE transactions on neural networks and learning systems},
  volume={32},
  number={1},
  pages={4--24},
  year={2020},
  publisher={IEEE}
}

@article{han2024retrieval,
  title={Retrieval-augmented generation with graphs (graphrag)},
  author={Han, Haoyu and Wang, Yu and Shomer, Harry and Guo, Kai and Ding, Jiayuan and Lei, Yongjia and Halappanavar, Mahantesh and Rossi, Ryan A and Mukherjee, Subhabrata and Tang, Xianfeng and others},
  journal={arXiv preprint arXiv:2501.00309},
  year={2024}
}

@article{peng2024graph,
  title={Graph retrieval-augmented generation: A survey},
  author={Peng, Boci and Zhu, Yun and Liu, Yongchao and Bo, Xiaohe and Shi, Haizhou and Hong, Chuntao and Zhang, Yan and Tang, Siliang},
  journal={arXiv preprint arXiv:2408.08921},
  year={2024}
}

@article{yasunaga2021qa,
  title={QA-GNN: Reasoning with language models and knowledge graphs for question answering},
  author={Yasunaga, Michihiro and Ren, Hongyu and Bosselut, Antoine and Liang, Percy and Leskovec, Jure},
  journal={arXiv preprint arXiv:2104.06378},
  year={2021}
}

@article{yasunaga2022deep,
  title={Deep bidirectional language-knowledge graph pretraining},
  author={Yasunaga, Michihiro and Bosselut, Antoine and Ren, Hongyu and Zhang, Xikun and Manning, Christopher D and Liang, Percy S and Leskovec, Jure},
  journal={Advances in Neural Information Processing Systems},
  volume={35},
  pages={37309--37323},
  year={2022}
}

@article{zhang2022greaselm,
  title={Greaselm: Graph reasoning enhanced language models for question answering},
  author={Zhang, Xikun and Bosselut, Antoine and Yasunaga, Michihiro and Ren, Hongyu and Liang, Percy and Manning, Christopher D and Leskovec, Jure},
  journal={arXiv preprint arXiv:2201.08860},
  year={2022}
}

@article{feng2020scalable,
  title={Scalable multi-hop relational reasoning for knowledge-aware question answering},
  author={Feng, Yanlin and Chen, Xinyue and Lin, Bill Yuchen and Wang, Peifeng and Yan, Jun and Ren, Xiang},
  journal={arXiv preprint arXiv:2005.00646},
  year={2020}
}

@article{mavromatis2024gnn,
  title={Gnn-rag: Graph neural retrieval for large language model reasoning},
  author={Mavromatis, Costas and Karypis, George},
  journal={arXiv preprint arXiv:2405.20139},
  year={2024}
}

@article{liu2024dual,
  title={Dual reasoning: A gnn-llm collaborative framework for knowledge graph question answering},
  author={Liu, Guangyi and Zhang, Yongqi and Li, Yong and Yao, Quanming},
  journal={arXiv preprint arXiv:2406.01145},
  year={2024}
}

@article{luo2023reasoning,
  title={Reasoning on graphs: Faithful and interpretable large language model reasoning},
  author={Luo, Linhao and Li, Yuan-Fang and Haffari, Gholamreza and Pan, Shirui},
  journal={arXiv preprint arXiv:2310.01061},
  year={2023}
}

@article{wu2023retrieve,
  title={Retrieve-rewrite-answer: A kg-to-text enhanced llms framework for knowledge graph question answering},
  author={Wu, Yike and Hu, Nan and Bi, Sheng and Qi, Guilin and Ren, Jie and Xie, Anhuan and Song, Wei},
  journal={arXiv preprint arXiv:2309.11206},
  year={2023}
}

@article{al2020named,
  title={Named entity extraction for knowledge graphs: A literature overview},
  author={Al-Moslmi, Tareq and Oca{\~n}a, Marc Gallofr{\'e} and Opdahl, Andreas L and Veres, Csaba},
  journal={IEEE access},
  volume={8},
  pages={32862--32881},
  year={2020},
  publisher={IEEE}
}

@article{wen2023mindmap,
  title={Mindmap: Knowledge graph prompting sparks graph of thoughts in large language models},
  author={Wen, Yilin and Wang, Zifeng and Sun, Jimeng},
  journal={arXiv preprint arXiv:2308.09729},
  year={2023}
}

@article{sanmartin2024kg,
  title={Kg-rag: Bridging the gap between knowledge and creativity},
  author={Sanmartin, Diego},
  journal={arXiv preprint arXiv:2405.12035},
  year={2024}
}

@article{choudhary2023complex,
  title={Complex logical reasoning over knowledge graphs using large language models},
  author={Choudhary, Nurendra and Reddy, Chandan K},
  journal={arXiv preprint arXiv:2305.01157},
  year={2023}
}

@article{he2024g,
  title={G-retriever: Retrieval-augmented generation for textual graph understanding and question answering},
  author={He, Xiaoxin and Tian, Yijun and Sun, Yifei and Chawla, Nitesh and Laurent, Thomas and LeCun, Yann and Bresson, Xavier and Hooi, Bryan},
  journal={Advances in Neural Information Processing Systems},
  volume={37},
  pages={132876--132907},
  year={2024}
}

@inproceedings{fang2024reano,
  title={Reano: Optimising retrieval-augmented reader models through knowledge graph generation},
  author={Fang, Jinyuan and Meng, Zaiqiao and Macdonald, Craig},
  booktitle={Proceedings of the 62nd Annual Meeting of the Association for Computational Linguistics (Volume 1: Long Papers)},
  pages={2094--2112},
  year={2024}
}

@book{ma2021deep,
  title={Deep learning on graphs},
  author={Ma, Yao and Tang, Jiliang},
  year={2021},
  publisher={Cambridge University Press}
}

@article{jiang2024kg,
  title={Kg-agent: An efficient autonomous agent framework for complex reasoning over knowledge graph},
  author={Jiang, Jinhao and Zhou, Kun and Zhao, Wayne Xin and Song, Yang and Zhu, Chen and Zhu, Hengshu and Wen, Ji-Rong},
  journal={arXiv preprint arXiv:2402.11163},
  year={2024}
}

@article{zhu2024knowagent,
  title={Knowagent: Knowledge-augmented planning for llm-based agents},
  author={Zhu, Yuqi and Qiao, Shuofei and Ou, Yixin and Deng, Shumin and Lyu, Shiwei and Shen, Yue and Liang, Lei and Gu, Jinjie and Chen, Huajun and Zhang, Ningyu},
  journal={arXiv preprint arXiv:2403.03101},
  year={2024}
}

@article{nezhurina2024alice,
  title={Alice in wonderland: Simple tasks showing complete reasoning breakdown in state-of-the-art large language models},
  author={Nezhurina, Marianna and Cipolina-Kun, Lucia and Cherti, Mehdi and Jitsev, Jenia},
  journal={arXiv preprint arXiv:2406.02061},
  year={2024}
}

@article{zheng2023large,
  title={Large language models are not robust multiple choice selectors},
  author={Zheng, Chujie and Zhou, Hao and Meng, Fandong and Zhou, Jie and Huang, Minlie},
  journal={arXiv preprint arXiv:2309.03882},
  year={2023}
}

@article{wei2022chain,
  title={Chain-of-thought prompting elicits reasoning in large language models},
  author={Wei, Jason and Wang, Xuezhi and Schuurmans, Dale and Bosma, Maarten and Xia, Fei and Chi, Ed and Le, Quoc V and Zhou, Denny and others},
  journal={Advances in neural information processing systems},
  volume={35},
  pages={24824--24837},
  year={2022}
}

@inproceedings{besta2024graph,
  title={Graph of thoughts: Solving elaborate problems with large language models},
  author={Besta, Maciej and Blach, Nils and Kubicek, Ales and Gerstenberger, Robert and Podstawski, Michal and Gianinazzi, Lukas and Gajda, Joanna and Lehmann, Tomasz and Niewiadomski, Hubert and Nyczyk, Piotr and others},
  booktitle={Proceedings of the AAAI conference on artificial intelligence},
  volume={38},
  number={16},
  pages={17682--17690},
  year={2024}
}

@article{deng2021reasonbert,
  title={ReasonBERT: Pre-trained to reason with distant supervision},
  author={Deng, Xiang and Su, Yu and Lees, Alyssa and Wu, You and Yu, Cong and Sun, Huan},
  journal={arXiv preprint arXiv:2109.04912},
  year={2021}
}

@article{lobo2024impact,
  title={On the impact of fine-tuning on chain-of-thought reasoning},
  author={Lobo, Elita and Agarwal, Chirag and Lakkaraju, Himabindu},
  journal={arXiv preprint arXiv:2411.15382},
  year={2024}
}

@article{xu2025towards,
  title={Towards large reasoning models: A survey of reinforced reasoning with large language models},
  author={Xu, Fengli and Hao, Qianyue and Zong, Zefang and Wang, Jingwei and Zhang, Yunke and Wang, Jingyi and Lan, Xiaochong and Gong, Jiahui and Ouyang, Tianjian and Meng, Fanjin and others},
  journal={arXiv preprint arXiv:2501.09686},
  year={2025}
}

@article{sui2025stop,
  title={Stop overthinking: A survey on efficient reasoning for large language models},
  author={Sui, Yang and Chuang, Yu-Neng and Wang, Guanchu and Zhang, Jiamu and Zhang, Tianyi and Yuan, Jiayi and Liu, Hongyi and Wen, Andrew and Zhong, Shaochen and Zou, Na and others},
  journal={arXiv preprint arXiv:2503.16419},
  year={2025}
}

@article{rafailov2023direct,
  title={Direct preference optimization: Your language model is secretly a reward model},
  author={Rafailov, Rafael and Sharma, Archit and Mitchell, Eric and Manning, Christopher D and Ermon, Stefano and Finn, Chelsea},
  journal={Advances in neural information processing systems},
  volume={36},
  pages={53728--53741},
  year={2023}
}

@article{zhang2024rest,
  title={Rest-mcts*: Llm self-training via process reward guided tree search},
  author={Zhang, Dan and Zhoubian, Sining and Hu, Ziniu and Yue, Yisong and Dong, Yuxiao and Tang, Jie},
  journal={Advances in Neural Information Processing Systems},
  volume={37},
  pages={64735--64772},
  year={2024}
}

@article{dai2024large,
  title={How do large language models understand graph patterns? a benchmark for graph pattern comprehension},
  author={Dai, Xinnan and Qu, Haohao and Shen, Yifen and Zhang, Bohang and Wen, Qihao and Fan, Wenqi and Li, Dongsheng and Tang, Jiliang and Shan, Caihua},
  journal={arXiv preprint arXiv:2410.05298},
  year={2024}
}

@article{kazemi2018simple,
  title={Simple embedding for link prediction in knowledge graphs},
  author={Kazemi, Seyed Mehran and Poole, David},
  journal={Advances in neural information processing systems},
  volume={31},
  year={2018}
}

@inproceedings{dutt2022perkgqa,
  title={PerKGQA: Question answering over personalized knowledge graphs},
  author={Dutt, Ritam and Bhattacharjee, Kasturi and Gangadharaiah, Rashmi and Roth, Dan and Rose, Carolyn},
  booktitle={Findings of the Association for Computational Linguistics: NAACL 2022},
  pages={253--268},
  year={2022}
}

@inproceedings{yih2016value,
  title={The value of semantic parse labeling for knowledge base question answering},
  author={Yih, Wen-tau and Richardson, Matthew and Meek, Christopher and Chang, Ming-Wei and Suh, Jina},
  booktitle={Proceedings of the 54th Annual Meeting of the Association for Computational Linguistics (Volume 2: Short Papers)},
  pages={201--206},
  year={2016}
}

@article{talmor2018web,
  title={The web as a knowledge-base for answering complex questions},
  author={Talmor, Alon and Berant, Jonathan},
  journal={arXiv preprint arXiv:1803.06643},
  year={2018}
}

@article{touvron2023llama,
  title={Llama: Open and efficient foundation language models},
  author={Touvron, Hugo and Lavril, Thibaut and Izacard, Gautier and Martinet, Xavier and Lachaux, Marie-Anne and Lacroix, Timoth{\'e}e and Rozi{\`e}re, Baptiste and Goyal, Naman and Hambro, Eric and Azhar, Faisal and others},
  journal={arXiv preprint arXiv:2302.13971},
  year={2023}
}

@inproceedings{bollacker2008freebase,
  title={Freebase: a collaboratively created graph database for structuring human knowledge},
  author={Bollacker, Kurt and Evans, Colin and Paritosh, Praveen and Sturge, Tim and Taylor, Jamie},
  booktitle={Proceedings of the 2008 ACM SIGMOD international conference on Management of data},
  pages={1247--1250},
  year={2008}
}

@article{wang2023keqing,
  title={keqing: knowledge-based question answering is a nature chain-of-thought mentor of LLM},
  author={Wang, Chaojie and Xu, Yishi and Peng, Zhong and Zhang, Chenxi and Chen, Bo and Wang, Xinrun and Feng, Lei and An, Bo},
  journal={arXiv preprint arXiv:2401.00426},
  year={2023}
}

@article{luo2025graph,
  title={Graph-r1: Towards agentic graphrag framework via end-to-end reinforcement learning},
  author={Luo, Haoran and Chen, Guanting and Lin, Qika and Guo, Yikai and Xu, Fangzhi and Kuang, Zemin and Song, Meina and Wu, Xiaobao and Zhu, Yifan and Tuan, Luu Anh and others},
  journal={arXiv preprint arXiv:2507.21892},
  year={2025}
}

@article{yu2025graphrag,
  title={GraphRAG-R1: Graph Retrieval-Augmented Generation with Process-Constrained Reinforcement Learning},
  author={Yu, Chuanyue and Zhao, Kuo and Li, Yuhan and Chang, Heng and Feng, Mingjian and Jiang, Xiangzhe and Sun, Yufei and Li, Jia and Zhang, Yuzhi and Li, Jianxin and others},
  journal={arXiv preprint arXiv:2507.23581},
  year={2025}
}

@article{zhang2022subgraph,
  title={Subgraph retrieval enhanced model for multi-hop knowledge base question answering},
  author={Zhang, Jing and Zhang, Xiaokang and Yu, Jifan and Tang, Jian and Tang, Jie and Li, Cuiping and Chen, Hong},
  journal={arXiv preprint arXiv:2202.13296},
  year={2022}
}

@article{sun2019pullnet,
  title={Pullnet: Open domain question answering with iterative retrieval on knowledge bases and text},
  author={Sun, Haitian and Bedrax-Weiss, Tania and Cohen, William W},
  journal={arXiv preprint arXiv:1904.09537},
  year={2019}
}

@article{guo2025deepseek,
  title={Deepseek-r1: Incentivizing reasoning capability in llms via reinforcement learning},
  author={Guo, Daya and Yang, Dejian and Zhang, Haowei and Song, Junxiao and Zhang, Ruoyu and Xu, Runxin and Zhu, Qihao and Ma, Shirong and Wang, Peiyi and Bi, Xiao and others},
  journal={arXiv preprint arXiv:2501.12948},
  year={2025}
}

@article{ouyang2022training,
  title={Training language models to follow instructions with human feedback},
  author={Ouyang, Long and Wu, Jeffrey and Jiang, Xu and Almeida, Diogo and Wainwright, Carroll and Mishkin, Pamela and Zhang, Chong and Agarwal, Sandhini and Slama, Katarina and Ray, Alex and others},
  journal={Advances in neural information processing systems},
  volume={35},
  pages={27730--27744},
  year={2022}
}

@article{schulman2017proximal,
  title={Proximal policy optimization algorithms},
  author={Schulman, John and Wolski, Filip and Dhariwal, Prafulla and Radford, Alec and Klimov, Oleg},
  journal={arXiv preprint arXiv:1707.06347},
  year={2017}
}

@inproceedings{zhang2017variational,
  title={Variational Reasoning for Question Answering with Knowledge Graph},
  author={Zhang, Yuyu and Dai, Hanjun and Kozareva, Zornitsa and Smola, Alexander J and Song, Le},
  booktitle={AAAI},
  year={2018}
}

@article{zhou2018interpretable,
  title={An interpretable reasoning network for multi-relation question answering},
  author={Zhou, Mantong and Huang, Minlie and Zhu, Xiaoyan},
  journal={arXiv preprint arXiv:1801.04726},
  year={2018}
}

@article{miller2016key,
  title={Key-value memory networks for directly reading documents},
  author={Miller, Alexander and Fisch, Adam and Dodge, Jesse and Karimi, Amir-Hossein and Bordes, Antoine and Weston, Jason},
  journal={arXiv preprint arXiv:1606.03126},
  year={2016}
}

@inproceedings{he2021improving,
  title={Improving multi-hop knowledge base question answering by learning intermediate supervision signals},
  author={He, Gaole and Lan, Yunshi and Jiang, Jing and Zhao, Wayne Xin and Wen, Ji-Rong},
  booktitle={Proceedings of the 14th ACM international conference on web search and data mining},
  pages={553--561},
  year={2021}
}

@article{guo2025empowering,
  title={Empowering graphrag with knowledge filtering and integration},
  author={Guo, Kai and Shomer, Harry and Zeng, Shenglai and Han, Haoyu and Wang, Yu and Tang, Jiliang},
  journal={arXiv preprint arXiv:2503.13804},
  year={2025}
}

@article{li2024simple,
  title={Simple is effective: The roles of graphs and large language models in knowledge-graph-based retrieval-augmented generation},
  author={Li, Mufei and Miao, Siqi and Li, Pan},
  journal={arXiv preprint arXiv:2410.20724},
  year={2024}
}

@article{chung2024scaling,
  title={Scaling instruction-finetuned language models},
  author={Chung, Hyung Won and Hou, Le and Longpre, Shayne and Zoph, Barret and Tay, Yi and Fedus, William and Li, Yunxuan and Wang, Xuezhi and Dehghani, Mostafa and Brahma, Siddhartha and others},
  journal={Journal of Machine Learning Research},
  volume={25},
  number={70},
  pages={1--53},
  year={2024}
}

@misc{taori2023stanford,
  title={Stanford alpaca: An instruction-following llama model},
  author={Taori, Rohan and Gulrajani, Ishaan and Zhang, Tianyi and Dubois, Yann and Li, Xuechen and Guestrin, Carlos and Liang, Percy and Hashimoto, Tatsunori B},
  year={2023},
  publisher={Stanford, CA, USA}
}

@article{sun2018open,
  title={Open domain question answering using early fusion of knowledge bases and text},
  author={Sun, Haitian and Dhingra, Bhuwan and Zaheer, Manzil and Mazaitis, Kathryn and Salakhutdinov, Ruslan and Cohen, William W},
  journal={arXiv preprint arXiv:1809.00782},
  year={2018}
}

@article{jiang2022unikgqa,
  title={Unikgqa: Unified retrieval and reasoning for solving multi-hop question answering over knowledge graph},
  author={Jiang, Jinhao and Zhou, Kun and Zhao, Wayne Xin and Wen, Ji-Rong},
  journal={arXiv preprint arXiv:2212.00959},
  year={2022}
}

@article{sun2023think,
  title={Think-on-graph: Deep and responsible reasoning of large language model on knowledge graph},
  author={Sun, Jiashuo and Xu, Chengjin and Tang, Lumingyuan and Wang, Saizhuo and Lin, Chen and Gong, Yeyun and Ni, Lionel M and Shum, Heung-Yeung and Guo, Jian},
  journal={arXiv preprint arXiv:2307.07697},
  year={2023}
}

@article{wang2023knowledge,
  title={Knowledge-driven cot: Exploring faithful reasoning in llms for knowledge-intensive question answering},
  author={Wang, Keheng and Duan, Feiyu and Wang, Sirui and Li, Peiguang and Xian, Yunsen and Yin, Chuantao and Rong, Wenge and Xiong, Zhang},
  journal={arXiv preprint arXiv:2308.13259},
  year={2023}
}

@article{achiam2023gpt,
  title={Gpt-4 technical report},
  author={Achiam, Josh and Adler, Steven and Agarwal, Sandhini and Ahmad, Lama and Akkaya, Ilge and Aleman, Florencia Leoni and Almeida, Diogo and Altenschmidt, Janko and Altman, Sam and Anadkat, Shyamal and others},
  journal={arXiv preprint arXiv:2303.08774},
  year={2023}
}

@article{hu2022lora,
  title={Lora: Low-rank adaptation of large language models.},
  author={Hu, Edward J and Shen, Yelong and Wallis, Phillip and Allen-Zhu, Zeyuan and Li, Yuanzhi and Wang, Shean and Wang, Lu and Chen, Weizhu and others},
  journal={ICLR},
  volume={1},
  number={2},
  pages={3},
  year={2022}
}

@inproceedings{wang2024knowledge,
  title={Knowledge graph prompting for multi-document question answering},
  author={Wang, Yu and Lipka, Nedim and Rossi, Ryan A and Siu, Alexa and Zhang, Ruiyi and Derr, Tyler},
  booktitle={Proceedings of the AAAI conference on artificial intelligence},
  volume={38},
  number={17},
  pages={19206--19214},
  year={2024}
}

@inproceedings{lei-etal-2025-mixture,
    title = "Mixture of Structural-and-Textual Retrieval over Text-rich Graph Knowledge Bases",
    author = "Lei, Yongjia  and
      Han, Haoyu  and
      Rossi, Ryan A.  and
      Dernoncourt, Franck  and
      Lipka, Nedim  and
      Halappanavar, Mahantesh M  and
      Tang, Jiliang  and
      Wang, Yu",
    booktitle = "Findings of the Association for Computational Linguistics: ACL 2025",
    month = jul,
    year = "2025",
    address = "Vienna, Austria",
    publisher = "Association for Computational Linguistics",
}

@article{lei2025mixture,
  title={Mixture of Structural-and-Textual Retrieval over Text-rich Graph Knowledge Bases},
  author={Lei, Yongjia and Han, Haoyu and Rossi, Ryan A and Dernoncourt, Franck and Lipka, Nedim and Halappanavar, Mahantesh M and Tang, Jiliang and Wang, Yu},
  journal={arXiv preprint arXiv:2502.20317},
  year={2025}
}

@article{han2025rag,
  title={Rag vs. graphrag: A systematic evaluation and key insights},
  author={Han, Haoyu and Ma, Li and Shomer, Harry and Wang, Yu and Lei, Yongjia and Guo, Kai and Hua, Zhigang and Long, Bo and Liu, Hui and Aggarwal, Charu C and others},
  journal={arXiv preprint arXiv:2502.11371},
  year={2025}
}

@article{han2025reasoning,
  title={Reasoning with graphs: Structuring implicit knowledge to enhance llms reasoning},
  author={Han, Haoyu and Xie, Yaochen and Liu, Hui and Tang, Xianfeng and Nag, Sreyashi and Headden, William and Li, Yang and Luo, Chen and Ji, Shuiwang and He, Qi and others},
  journal={arXiv preprint arXiv:2501.07845},
  year={2025}
}

@article{yao2023tree,
  title={Tree of thoughts: Deliberate problem solving with large language models},
  author={Yao, Shunyu and Yu, Dian and Zhao, Jeffrey and Shafran, Izhak and Griffiths, Tom and Cao, Yuan and Narasimhan, Karthik},
  journal={Advances in neural information processing systems},
  volume={36},
  pages={11809--11822},
  year={2023}
}

\appendix
\clearpage
\balance

\section{Algorithm of SFT Dataset Construction}
\label{sec:appendix:a}

In this section, we detail the algorithm to construct the SFT datasets to train \method. The algorithm is summarized in Algorithm~\ref{alg:mine_and_sft}, where lines 1–14 construct all gold-consistent reasoning paths up to length $L_{\max}$; line 20 derives the step-wise gold answers; and lines 21–22 generate the step-wise new exploration paths. 
\section{Datasets}
In the experiments, we adopt four different datasets, i.e. \textbf{WebQSP}~\cite{yih2016value} and \textbf{CWQ}~\cite{talmor2018web}, \textbf{MetaQA}~\cite{zhang2017variational} and \textbf{PathQuestion}~\cite{zhou2018interpretable}. For the WebQSP and CWQ dataset, we leverage the processed dataset in \citet{luo2023reasoning}. For the PathQuestion and MetaQA dataset, we randomly sample 1,000 samples of two hop questions. The statistics of these datasets are shown in Table~\ref{tab:datasets}.

\begin{table}[H]
\centering
\caption{Statistics of datasets.}
\label{tab:datasets}
\vspace{-0.1in}
\begin{tabular}{l|c|c|c}
\hline
\textbf{Datasets} & \textbf{\#Train} & \textbf{\#Test} & \textbf{Max \#hop} \\
\hline
WebQSP & 2,826 & 1,628 & 2 \\
CWQ    & 27,639 & 3,531 & 4 \\
PathQuestion & 0 & 1,000 & 2 \\
MetaQA & 0 & 1,000 & 2 \\
\hline
\end{tabular}
\end{table}

\section{Experimental Settings}
In this section, we present the detailed experimental settings for \method.  
For \method\ and most of baseline models involving LLMs, we use the \textbf{Llama-3.1-8B-Instruct} model~\cite{touvron2023llama} as the backbone.  
We first leverage Algorithm~\ref{alg:mine_and_sft} to construct the gold trajectories for supervised fine-tuning.  
The training data of both WebQSP and CWQ are divided into two parts: 60\% for the \textbf{Supervised Fine-Tuning (SFT)} stage and the remaining 40\% for the \textbf{Reinforcement Learning (RL)} stage.  
To reduce memory consumption, we adopt \textbf{LoRA}~\cite{hu2022lora} in both stages with a rank of 32.  

During SFT, the loss is calculated only on the action prediction component, and the model is optimized using the AdamW optimizer with a learning rate of $1\times10^{-4}$.  
For the RL stage, we apply a smaller learning rate of $5\times10^{-6}$ and use the Group Relative Policy Optimization (GRPO) algorithm to optimize the policy with rule-based rewards. For each sample, we generate 4 responses.
Each training step uses a batch size of 4 with gradient accumulation to fit within GPU memory.  
Training is performed on 2$\times$H200 GPUs.

During inference, we follow a depth-first exploration strategy.  
Because each node in the knowledge graph may have numerous neighbors that cannot fit into the LLM context window at once, 
we split the neighbors into multiple batches and iteratively feed them to the model.  
We use a depth-first search algorithm to explore the graph, and we set the maximum exploration depth to 5 hops to prevent overly long reasoning chains and context overflow.  
For reproducibility, all random seeds are fixed, and the same prompt templates for SFT and RL.

\section{Results with different backbone models}
\label{app:sec:backbone}
To evaluate robustness across model sizes, we additionally train RoE-3B and other baseline models with Llama-3.2-3B-Instruct backbone. All models are trained on WebQSP and evaluated on both WebQSP and MetaQA.

\begin{algorithm}[H]
\caption{\textbf{SFT Dataset Construction}}
\label{alg:mine_and_sft}
\begin{algorithmic}[1]
\Require Question $q$; KG $\mathcal{G}=\{\mathcal{V},\mathcal{R},\mathcal{E}\}$; seed entities $\mathcal{V}_0$; gold answers $\mathcal{A}^*$; max path length $L_{\max}$
\Ensure Dataset $\mathcal{D}_{\text{SFT}}=\{(s_d,\ a_d^{*})\}_{d=0}^{L_{\max}-1}$ with $a_d^{*}=(\mathcal{A}_d^{*}, \mathcal{P}_d^{*})$

\State \textbf{Phase I: Mine all gold-consistent paths (length $\le L_{\max}$)}
\State $\Pi \gets \emptyset$;\quad $\mathsf{Q} \gets \{[v_0] \mid v_0 \in \mathcal{V}_0\}$
\While{$\mathsf{Q}$ not empty}
  \State $p \gets$ pop from $\mathsf{Q}$;\quad $v \gets \text{frontier}(p)$
  \If{$|p|-1 > L_{\max}$} 
  \textbf{continue} 
  \EndIf
  \If{$v \in \mathcal{A}^*$} \ $\Pi \gets \Pi \cup \{p\}$ \EndIf
  \ForAll{$(v,r,u)\in\mathcal{E}$}
     \If{$u \notin p$} \Comment{simple-path constraint (avoid cycles)}
        \State $\mathsf{Q} \gets \mathsf{Q} \cup \{\, p \oplus (v,r,u) \,\}$
     \EndIf
  \EndFor
\EndWhile

\Statex \textbf{Phase II: Build step-wise actions from mined paths}
\State Define depth-indexed prefix pools for $d=0,\dots,L_{\max}$:
\[
\mathsf{Pref}(d) \ = \ \{\ \text{all prefixes of any } p\in\Pi \text{ with } |{\rm prefix}|-1=d\ \}
\]
\State $\mathcal{D}_{\text{SFT}} \gets \emptyset$
\For{$d=0$ \textbf{to} $L_{\max}-1$}
   \State $\mathcal{P}_d \gets \mathsf{Pref}(d)$;\quad $\mathcal{F}_d \gets \{\text{frontier}(p)\mid p\in\mathcal{P}_d\}$
   \State \textbf{State:}\quad $s_d \gets (q,\ \mathcal{P}_d,\ \mathcal{N}_{\mathcal{F}_d})$

\State \textbf{Gold step-answers:}
   \[
   \mathcal{A}_d^{*}\ =\ \{\, u \in \mathcal{A}^* \ \mid\ \exists (v,r,u)\in\mathcal{N}_{\mathcal{F}_d} \,\}
   \]

   \State \textbf{Gold one-hop extensions:}
   \[
   \mathcal{E}^{\Pi}_d\ =\ \{\, (p,(v,r,u)) \in \mathsf{Pref}(d+1)\ \ \mid\ p\in\mathcal{P}_d,\ \text{frontier}(p)=v \}
   \]

   \State \textbf{Same-relation sibling expansion:}
   \[
   \mathcal{P}_d^{*}\ =\ \big\{\, p \oplus (v,r,u') \ \big|\ (p,(v,r,u))\in \mathcal{E}^{\Pi}_d,\ (v,r,u')\in\mathcal{E} \big\}
   \]

   \State \textbf{Record step:}\ \ $a_d^{*}\gets(\mathcal{A}_d^{*}, \mathcal{P}_d^{*})$;\ \ 
          $\mathcal{D}_{\text{SFT}} \gets \mathcal{D}_{\text{SFT}} \cup \{(s_d,\ a_d^{*})\}$
\EndFor

\State \Return $\mathcal{D}_{\text{SFT}}$
\end{algorithmic}
\end{algorithm}

\begin{table}[htb]
\centering
\caption{Performance comparison on the WebQSP and MetaQA datasets. All models use the Llama-3.2-3B-Instruct backbone and are trained on the WebQSP dataset.}
\label{app:table:3B}
\resizebox{\linewidth}{!}{%
\begin{tabular}{c|cc|cc}
\hline
 & \multicolumn{2}{c|}{WebQSP $\rightarrow$ WebQSP} & \multicolumn{2}{c}{WebQSP $\rightarrow$ MetaQA} \\ \hline
Method & Hit & F1 & Hit & F1 \\ \hline
G-Retrieval-3B & 67.32 & 47.25 & 12.7 & 5.58 \\
RoG-3B & 73.95 & 55.79 & 10.80 & 6.00 \\
GNN-RAG-3B & 81.51 & 67.16 & 41.70 & 13.94 \\
RoE-3B & \textbf{85.56} & \textbf{70.62} & \textbf{74.20} & \textbf{57.12} \\ \hline
\end{tabular}
}
\end{table}

From the results in Table~\ref{app:table:3B}, RoE continues to outperform all baselines by a large margin, especially in the transfer setting on MetaQA. Notably, RoE-3B surpasses baselines even when they use Llama-8B, especially on the transfer setting. GRPO vs. GSPO yields similar performance, showing that RoE is not sensitive to the RL algorithm choice.

\end{document}